\documentclass[twocolumn,twocolappendix]{aastex701}

\usepackage{tabularx} %
\usepackage{amsmath,amssymb}  %
\usepackage{graphicx} %

\usepackage{textcomp, gensymb, bm}
\usepackage{caption}
\usepackage{subcaption}
\usepackage{breqn}
\usepackage{ulem}
\usepackage{soul}

\begin{document}

\title{Modern tidal interaction models for rapid binary population synthesis: II. Binary black hole formation, mergers, and spins}

\author[orcid=0000-0001-7344-5260,gname=Veome,sname=Kapil]{Veome Kapil}
\affiliation{William H. Miller III Department of Physics and Astronomy, Johns Hopkins University, \\3400 N. Charles Street, Baltimore, Maryland, 21218, USA}
\email{vkapil1@jh.edu}  

\author[orcid=0000-0002-6134-8946,gname=Ilya,sname=Mandel]{Ilya Mandel}
\affiliation{Monash Centre for Astrophysics, School of Physics and Astronomy, Monash University, Clayton, Victoria 3800, Australia}
\affiliation{OzGrav: Australian Research Council Centre of Excellence for Gravitational Wave Discovery, Clayton, VIC 3800, Australia}
\email{ilya.mandel@monash.edu}

\author[orcid=0000-0003-1530-2557,gname=Jeff,sname=Riley]{Jeff Riley}
\affiliation{Monash Centre for Astrophysics, School of Physics and Astronomy, Monash University, Clayton, Victoria 3800, Australia}
\affiliation{OzGrav: Australian Research Council Centre of Excellence for Gravitational Wave Discovery, Clayton, VIC 3800, Australia}
\email{jeff.riley@monash.edu}

\author[orcid=0000-0001-7113-723X,gname=Evgeni,sname=Grishin]{Evgeni Grishin}
\affiliation{Monash Centre for Astrophysics, School of Physics and Astronomy, Monash University, Clayton, Victoria 3800, Australia}
\affiliation{OzGrav: Australian Research Council Centre of Excellence for Gravitational Wave Discovery, Clayton, VIC 3800, Australia}
\email{evgeni.grishin@monash.edu}

\author[orcid=0000-0002-4544-0750,gname=Jim,sname=Fuller]{Jim Fuller}
\affiliation{TAPIR, Mailcode 350-17, California Institute of Technology, Pasadena, CA 91125, USA}
\email{jfuller@caltech.edu}

\author[orcid=0000-0003-0751-5130,gname=Emanuele,sname=Berti]{Emanuele Berti}
\affiliation{William H. Miller III Department of Physics and Astronomy, Johns Hopkins University, \\3400 N. Charles Street, Baltimore, Maryland, 21218, USA}
\email{berti@jhu.edu}

\begin{abstract}
We present predictions for the merger rates and effective spin ($\chi_{\rm eff}$) distribution of binary black holes (BBHs) from isolated binary evolution, using a new self-consistent tidal dissipation implementation in the rapid binary population synthesis code COMPAS. Most of the first-born black holes (BHs) in our simulated merging BBHs are formed with zero spins, with the exception of BBHs formed from chemically homogeneous evolution. The spins of the second-born BHs with the new model depend significantly on the efficiency of tidal dissipation and mass transfer history, and crucially, are not always consistent with pre-supernova synchronization. High-$\chi_{\rm eff}$ binaries preferentially merge at high redshift due to smaller binary separations at BBH formation and shorter coalescence times, thus rendering them largely inaccessible to current gravitational wave (GW) detectors. We expect the intrinsic spin distribution of merging BBHs formed from isolated evolution to be strongly biased toward low $\chi_{\rm eff}$ with current detectors, with a third of systems having $\chi_{\rm eff} < 0.05$ and only $\sim 3\%$ with $\chi_{\rm eff}>0.5$.  However, $\chi_{\rm eff}$ will increase as GW detectors become sensitive to higher redshift sources, with up to $\sim 15\%$ of systems having $\chi_{\rm eff}>0.5$.

\end{abstract}


\section{Introduction}
Gravitational wave (GW) observations have opened an unprecedented window into the properties of compact objects such as black holes (BHs) and neutron stars. Alongside measurements of masses and orbital parameters, the spins of compact objects, particularly the effective spin parameter $\chi_{\rm eff}$, are now being directly probed by current detectors including LIGO, Virgo, and KAGRA \citep{LIGOScientific:2018mvr, LIGOScientific:2020ibl, KAGRA:2021vkt, LIGOScientific:2025slb, LIGOScientific:2026wfs}. These have allowed for population-level inferences on the spins of double compact objects (DCOs) \citep[e.g.,][]{KAGRA:2021duu, Callister:2022qwb, LIGOScientific:2025pvj, LIGOScientific:2026ctl, Alvarez-Lopez:2026ymo}. Further robust measurements of individual binary spin magnitudes and orientations are anticipated with upcoming third-generation (3G) GW observatories such as the Einstein Telescope (ET) \citep{Punturo:2010zz, Hild:2010id,ET:2019dnz,ET:2025xjr} and Cosmic Explorer (CE) \citep{Reitze:2019iox, Evans:2021gyd}. These measurements should help better constrain astrophysical models and disentangle the evolutionary pathways that give rise to observed spin distributions, such as isolated binary evolution or dynamical assembly.

Isolated binary evolution is likely to contribute significantly to the observed population of compact binaries \citep[e.g.,][]{Dominik:2012kk, Dominik:2013tma, Dominik:2014yma, Broekgaarden:2021efa, Mandel:2018hfr}. In isolated evolution, the natal spin of a compact remnant depends sensitively on the angular momentum retained by the stellar core prior to collapse. BHs formed from rapidly rotating chemically homogeneous stars, or stripped stars in tight binaries, may yield high dimensionless spins \citep{deMink:2010zm, Zaldarriaga:2017qkw, Qin:2018sxk, Bavera:2020inc}, while efficient angular momentum transport or envelope stripping in close binaries can result in slowly spinning remnants \citep{Qin:2018vaa, Fuller:2019sxi}. Tidal interactions play a central role in regulating the spins of massive stellar binaries prior to core collapse and may imprint observable signatures in the spin magnitudes and alignments of compact binaries formed in isolation \citep{Gerosa:2013laa, Gerosa:2014kta, Wysocki:2017isg, Gerosa:2018wbw, Belczynski:2017gds, Steinle:2020xej, Baibhav:2020xdf, Gangardt:2021lic, Varma:2021xbh, Steinle:2022rhj, Singh:2025xns, Biscoveanu:2025jpc, Biscoveanu:2026ikx}. Furthermore, the rates of systems expected to host rapidly spinning compact objects, such as X-ray binaries and long gamma-ray bursts, are also sensitive to the efficiency of tidal spin-up and angular momentum retention \citep{Woosley:2005gy, Fragos:2014cva, Qin:2018sxk, Batta:2019clm}. 

Spin distributions and merger rates of DCOs are often produced with the aid of rapid population synthesis codes, which invoke several simplifying astrophysical assumptions to feasibly simulate large populations of isolated stars \citep[e.g.,][]{Bavera:2020inc, Zevin:2022wrw, Korb:2024igp, Olejak:2024qxr}. A common population synthesis approach to spin modeling is to evolve the binary orbit and stellar structure forward in time, and then to apply tidal and spin prescriptions as a post-processing step. While computationally tractable, this decoupling of tidal synchronization from orbital evolution means that the back-reaction of spin-orbit coupling on the binary separation, mass transfer episodes, and subsequent evolution is not captured self-consistently. Other studies  rely on the tidal formalisms of \citet{zahn_tidal_1977, hut_tidal_1981, Hurley:2002rf}, which do not account for detailed stellar interiors, internal stratification, or frequency-dependent viscosity as modeled in recent detailed simulations (see \citet{Kushnir:2017, Barker:2020MNRAS.498.2270B, Ahuir:2021} for updated descriptions of tidal dissipation in stars). There remains an opportunity to predict remnant spins using updated tidal models within a self-consistent binary evolution framework.

In this work, we present an application of our updated tidal prescription, introduced in our companion paper \citep{Kapil:2026hiw}, to simulations of DCOs in the COMPAS population synthesis code \citep{Stevenson:2017tfq, Vigna-Gomez:2018dza, COMPAS_2022, COMPAS_2025}. We briefly summarize the new model in Sec.~\ref{sec:tidal_models}, and we also introduce a toy model with instantaneous tidal circularization and synchronization to bracket the upper limits of tidal efficiency. In Sec.~\ref{sec:population_details}, we introduce the population parameters which are simulated using COMPAS. We apply our framework to study the formation and merger yields of binary black holes (BBHs) under various tidal prescriptions in Sec.~\ref{sec:bbh_merger_rates}, and discuss the resulting spins of BBHs in Sec.~\ref{sec:spin_magnitudes}. In Sec.~\ref{sec:bbh_astro_population} we re-weight our simulations based on metallicity-specific star formation history to present realistic distributions of BBH spins across multiple tidal prescriptions. Finally, we briefly explore the prospects of detecting BBH spins with current and future GW detectors in Sec.~\ref{sec:gw_detections}, and we summarize our results in Sec.~\ref{sec:conclusions}.

\section{Tidal dissipation models}
\label{sec:tidal_models}

\subsection{KAPIL26 Tides Model}
The fiducial model for tides used in this work comes from our companion paper \citep{Kapil:2026hiw}, which we henceforth abbreviate as KAPIL26. This prescription adapts modern theoretical predictions for both equilibrium and dynamical tidal dissipation, calibrated against detailed stellar models across the main sequence (MS), Hertzsprung gap (HG), and giant phases, to a semi-analytic population synthesis framework, allowing tidal torques to be applied self-consistently at each timestep rather than as a post-processing correction. We refer the reader to the companion paper for technical details; here we highlight the caveats most relevant to population-level predictions of compact object spins. 

First, the KAPIL26 tidal prescription is formally valid only in the quasi-circular limit ($e\ll 1$), even though isolated binary systems may exist across a broad range of eccentricities \citep{Duquennoy:1991zu, Raghavan:2010ApJS..190....1R, Moe:2017ApJS..230...15M}. Additionally, binaries may acquire significant eccentricities from supernova (SN) kicks during their evolution \citep{Blaauw:1961BAN....15..265B, Brandt:1994rr, Kalogera:1996rm, Gerosa:2013laa}. In this work, this limitation is largely mitigated by the assumed initial eccentricity distribution of our population (see Sec.~\ref{sec:population_details}) and the assumption of rapid binary circularization at the onset of Roche-lobe overflow (RLOF) in COMPAS. Tidal torques scale roughly as $(R_*/a)^6$, where $R_*$ is the stellar radius and $a$ is the semi-major axis of the binary, and binaries of interest to us undergo at least one RLOF episode. Binary separations are typically too large prior to RLOF for tides to be efficient, and mass transfer leads to circularization before tides can play a significant role in binary evolution. Therefore, most of the tidal dissipation occurs in the circular regime in our models. 

Second, we assume for the sake of tidal computations that stellar spin axes are aligned with the orbital angular momentum vector, i.e., zero spin-orbit misalignment for a given star until supernova. In COMPAS, supernova kicks can impart both a recoil velocity and a tilt to the orbital plane, inducing misalignment between the spin of the surviving companion and the new orbital axis \citep{Blaauw:1961BAN....15..265B, Brandt:1994rr, Kalogera:1999tq}. Tidal torques in misaligned systems drive the spin axis toward alignment on a timescale that depends on the degree of misalignment and the tidal dissipation efficiency \citep{hut_tidal_1981, Lai:2011vr}. However, incorporating three-dimensional spin-orbit dynamics for tides requires tracking the full spin vector for each star, which is not currently implemented in COMPAS. The aligned spin assumption is most consequential for the second-born compact object, whose progenitor may have experienced spin-orbit misalignment following the first supernova. For the purposes of this work, we note that tidal realignment is expected to be efficient for synchronized close binaries \citep{hut_tidal_1981, Lai:2011vr}, which are precisely the systems most likely to produce BBH mergers. A more careful treatment of tidal dissipation in misaligned binaries is left to future work.

Third, both COMPAS and the KAPIL26 prescription assume rigid-body rotation for each star, meaning the entire star is assigned a single spin rate at every timestep. In reality, stellar interiors can sustain significant differential rotation, wherein the core rotates considerably faster than the envelope depending on the coupling of angular momentum (AM) transport between layers (although AM transport may be efficient enough to support rigid-body rotation, see \citet{Fuller:2019ckz, Fuller:2019sxi}). This internal angular momentum distribution sets the spin of the resulting BH in the case of incomplete fallback after supernova. Improving upon the rigid-rotation assumption requires a detailed treatment of internal angular momentum transport, which remains considerably more uncertain than the tidal prescriptions themselves. In the context of our models, the assumption of rigid body rotation implies efficient angular momentum transport from the core to the outer layers of a star, which may be a reasonable assumption given recent theoretical estimates~\citep{Fuller:2019ckz}. We therefore note that our predicted BH spin magnitudes may carry an additional systematic uncertainty from this assumption, and leave a more self-consistent treatment of differential rotation to future work.

Finally, the accuracy of the KAPIL26 prescription for high-mass stars depends on characterizing the evolving boundary between convective and radiative regions, which sets the dominant tidal dissipation channel at a given evolutionary stage. In a rapid population synthesis framework like COMPAS, stellar structures are approximated from fitting formulae, meaning that transitions between convective and radiative envelopes as well as their respective tidal dissipation processes are also approximate. We refer the reader to KAPIL26 for comparisons to detailed stellar evolution, but we note that our models obtain the correct order of magnitude behavior across stellar types.

\subsection{PERFECT tides model}
\label{sec:perfect_tides_model}
As an upper limit on tidal interaction strength in binaries, we introduce a perfectly efficient tidal mechanism (hereafter, PERFECT tides) that can circularize and synchronize a given binary immediately, independently of stellar structure. This clearly unrealistic toy model provides a useful benchmark to compare against the more physically motivated KAPIL26 model.

Before resolving tidal interaction during each step of binary evolution, we can express the total angular momentum as a sum of the orbital and spin angular momenta, such that 
\begin{align}
    L_i &= I_{1} \Omega_{1, i} + I_{2} \Omega_{2, i} + L_{{\rm orb}, i}\,.
\end{align}
Here, $I_{n}$ is the moment of inertia of body $n$, $\Omega_{n, i}$ is its initial rotational velocity before the action of tides, $L_{{\rm orb}, i}$ is the initial orbital angular momentum of the binary before the action of tides, and $L_i$ is the initial total angular momentum.  Because we assume the angular momenta are all aligned when applying tides, it is sufficient to express the angular momenta and rotation vectors as scalars. The initial orbital angular momentum of the binary is given by 
\begin{equation}
    L_{{\rm orb}, i} = \sqrt{\frac{G M^2_{1} M^2_{2}}{(M_{1} + M_{2})} a_i (1-e^2_i)}\,.
\end{equation}
Here, $a_i$ is the initial semi-major axis, $e_i$ is the initial eccentricity, $G$ is the gravitational constant, and the component masses are labeled by $M_{n}$. As per the tidal model, we now impose that the binary becomes synchronized and circularized, with final angular momentum given by
\begin{align}
L_f &= I_{1} \Omega_{f} + I_{2} \Omega_{f} + L_{{\rm orb}, f}\,. 
\end{align}
Note that we have replaced the separate stellar rotational velocities in favor of a single $\Omega_f$, which encodes our synchronization condition. $L_{{\rm orb}, f}$ can be calculated for a circular binary as
\begin{align}
    L_{{\rm orb}, f} = G^{2/3} \frac{M_{1} M_{2}}{(M_{1} + M_{2})^{1/3}}  \Omega_{f}^{-1/3}\,.
\end{align}
Equating the initial and final angular momenta within the tidal evolution time step, $L_f = L_i$, we get
\begin{multline}
    I_{1} \Omega_{1, i} + I_{2} \Omega_{2, i} + \sqrt{\frac{G M^2_{1} M^2_{2}}{(M_{1} + M_{2})} a_i (1-e^2_i)} \\ = (I_{1} + I_{2}) \Omega_f + G^{2/3} \frac{M_{1} M_{2}}{(M_{1} + M_{2})^{1/3}}  \Omega_{f}^{-1/3}.
\end{multline}
This can be written as a quartic equation for $\Omega_f^{1/3}$:
\begin{equation}
    A \Omega_f  + B \Omega_{f}^{-1/3} + C = 0\,,
\end{equation}
where
\begin{align}
    A &\equiv I_{1} + I_{2}\,,\\
    B &\equiv G^{2/3} \frac{M_{1} M_{2}}{(M_{1} + M_{2})^{1/3}}\,, \\
    C &\equiv - \left(I_{1} \Omega_{1, i} + I_{2} \Omega_{2, i} + \sqrt{\frac{G M^2_{1} M^2_{2}}{(M_{1} + M_{2})} a_i (1-e^2_i)} \right)\,.
\end{align}
Once we solve for $\Omega_f$ using a root finder, we obtain $a_f$ using the Keplerian relation
\begin{equation}
    a_f^3 = \frac{G (M_{1} + M_{2})} {\Omega^{2}_f}\,.
\end{equation}
In the COMPAS implementation of the PERFECT model, we enforce synchronization and circularization across all stellar types, including binaries that may contain one or more compact objects. However, the negligibly small moments of inertia of compact objects mean that tidal coupling to a compact remnant has essentially no effect on the binary's orbital evolution.

\section{Population details}
\label{sec:population_details}
We use the COMPAS rapid binary population synthesis code to evolve three different populations of isolated binary stars. The first model is default COMPAS  without any tidal dissipation (except for chemically homogeneous stars), which is hereafter referred to as LEGACY. The second is the KAPIL26 prescription, and the third is the PERFECT tides prescription as defined in Sec.~\ref{sec:perfect_tides_model}. We simulate a total of $3.8 \times 10^6$ binary systems, with initial masses being drawn from the \cite{Kroupa:2000iv} initial mass function between $[5, 150] M_\odot$, and initial orbital periods being drawn uniformly in log space in the range $[1, 1000]$~days. The initial eccentricities are drawn from a power law $p(e)~\propto~e^{-0.42}$, which follows observations of binaries in young open clusters~\citep{Sana:2012px, deMink:2015yea}.  We sample metallicities between $Z=10^{-4}$ and $Z=0.03$ with a log-uniform metallicity distribution.  The rest of the initial conditions are set to their default values, and we use COMPAS v3.29.00 for all simulations shown in this paper.

For each tidal dissipation model considered in this work, we obtain BBH samples across a range of masses, separations, and metallicities. To determine potential GW sources, we select only binaries that merge within $\leq 13.8$~Gyr of formation. These represent the merging BBH samples from our simulations, although they may not be compatible with star formation history in the Universe or be detected with GW observatories. In the following two sections, we consider all such merging BBHs that emerge from our simulations regardless of formation or detectability likelihoods; we account for star formation rates in Sec.~\ref{sec:bbh_astro_population} and discuss observational selection effects in Sec.~\ref{sec:gw_detections}.

\section{Impact of tides on BBH merger yields}
\label{sec:bbh_merger_rates}
\begin{figure}
    \centering
    \includegraphics[width=\linewidth]{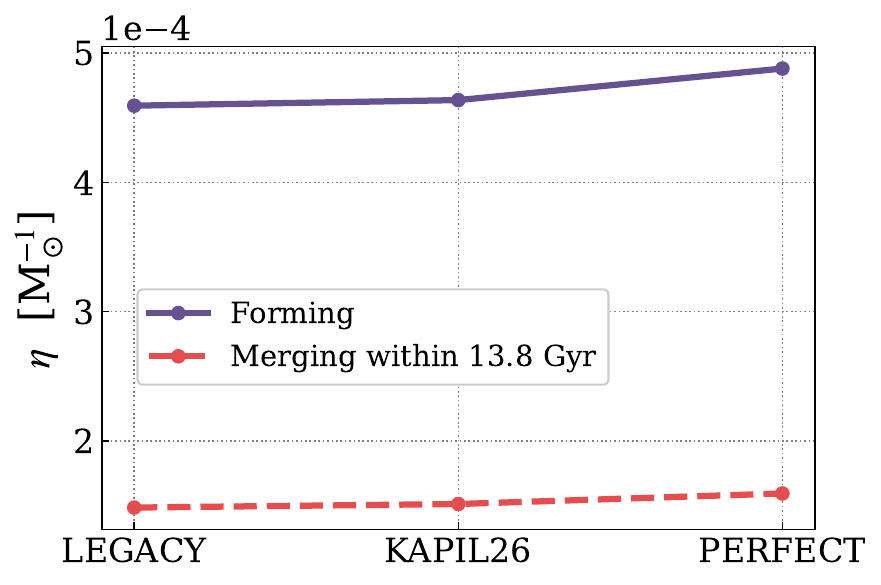}
    \caption{Yield of BBH formations and mergers, normalized to total star forming mass, across the three tidal prescriptions considered in this work. This figure integrates over the log-uniform distribution of simulated metallicities.}
    \label{fig:dco_rates}
\end{figure}

Varying the tidal dissipation strength in our simulations has a small but statistically notable effect on the overall formation and merger rates of BBHs. To demonstrate this, we compute the yield of BBHs as the ratio of the number of (merging) BBHs formed per unit star forming mass $M_{\rm SF}$,
\begin{equation}
    \eta = \frac{N_{\rm BBH}}{M_{\rm SF}} = \frac{N_{\rm BBH}  \cdot f_{\rm bin}}{ M_{\rm evolved}} \cdot \frac{\int_{5 M_\odot}^{150 M_\odot} m \cdot \xi(m) \; \mathrm{d}m}{\int_{0.08 M_\odot}^{150 M_\odot} m \cdot \xi(m) \; \mathrm{d}m}\,.
\end{equation}
In the above equation, $M_{\rm evolved}$ represents the total stellar mass we simulated while sampling the IMF $\xi(m)$ in the domain $[5, 150] M_\odot,$ which must be corrected to account for the full IMF range of stars in the Universe; $f_{\rm bin}$ is the fraction of stellar mass found in binaries, which can be well approximated as 1 for massive stars \citep{2017IAUS..329..110S}. In Fig.~\ref{fig:dco_rates} we show the total BBH yield per star forming mass, as well as the yield of BBHs that merge within 13.8 Gyr.

\begin{figure}
    \centering
    \includegraphics[width=\linewidth]{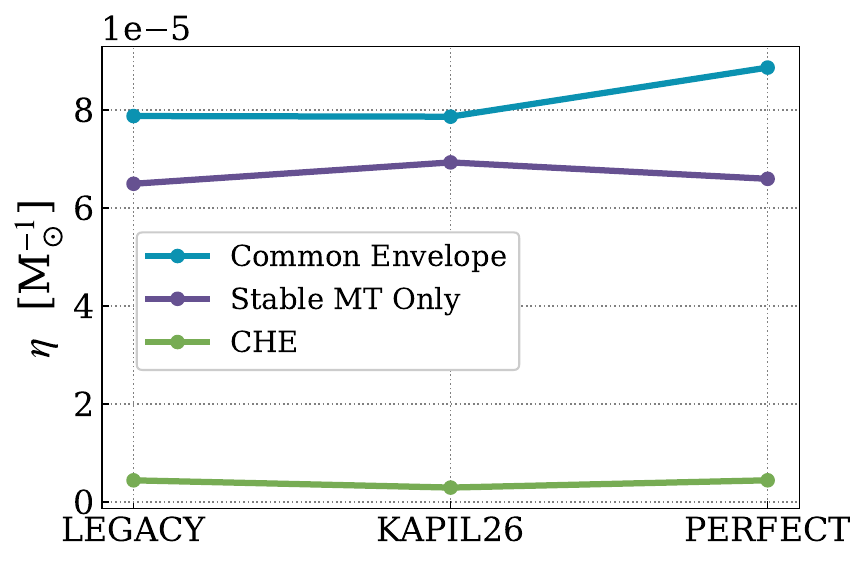}
    \caption{Yield of BBHs that merge within 13.8 Gyr by formation scenario, integrated over the log-uniform distribution of simulated metallicities, across the three tidal prescriptions considered in this work.
    }
    \label{fig:branching_ratios_mergers}
\end{figure}

In the broad context of binary evolution, we can think of our three simulations as representing choices on a continuous axis of tidal dissipation effectiveness, going from weakest (LEGACY) to strongest (PERFECT) tides. We may expect the formation and merger rates to vary monotonically with increasing tidal strength. This is generally true for the trends shown in Fig.~\ref{fig:dco_rates}. To better understand the various binary interactions that impact BBH rates, we show the yield of BBH mergers per formation pathway in Fig. \ref{fig:branching_ratios_mergers}. Here, we separate the BBH yield rates by binaries that experienced at least one common envelope (CE) phase, those that only experienced stable mass transfer (SMT), and those that underwent chemically homogeneous evolution (CHE).

\subsection{Common envelope}
\label{sec:ce_mergers}

All the progenitors of merging BBHs in our simulations experience at least one episode of mass transfer, wherein the donor overflows its Roche lobe and transfers mass and angular momentum to its companion. When mass transfer becomes dynamically unstable, the binary becomes engulfed in the envelope of the donor star. This unstable episode of mass transfer is known as a CE event. During CE, the binary experiences dynamical friction which significantly reduces the orbital separation, leading to very tight binaries that can merge efficiently. By default, COMPAS uses the $\alpha-\lambda$ prescription \citep{Webbink:1984ApJ...277..355W,Xu:2010wx} for resolving CEs. Across our simulations, approximately 92\% of BBHs formed via the CE channel end up merging within 13.8 Gyr, and it is the dominant contributing channel for BBH mergers from isolated binary evolution. 

\begin{figure*}
    \centering
    \begin{subfigure}[t]{0.49\textwidth}
        \centering
        \includegraphics[height=6in]{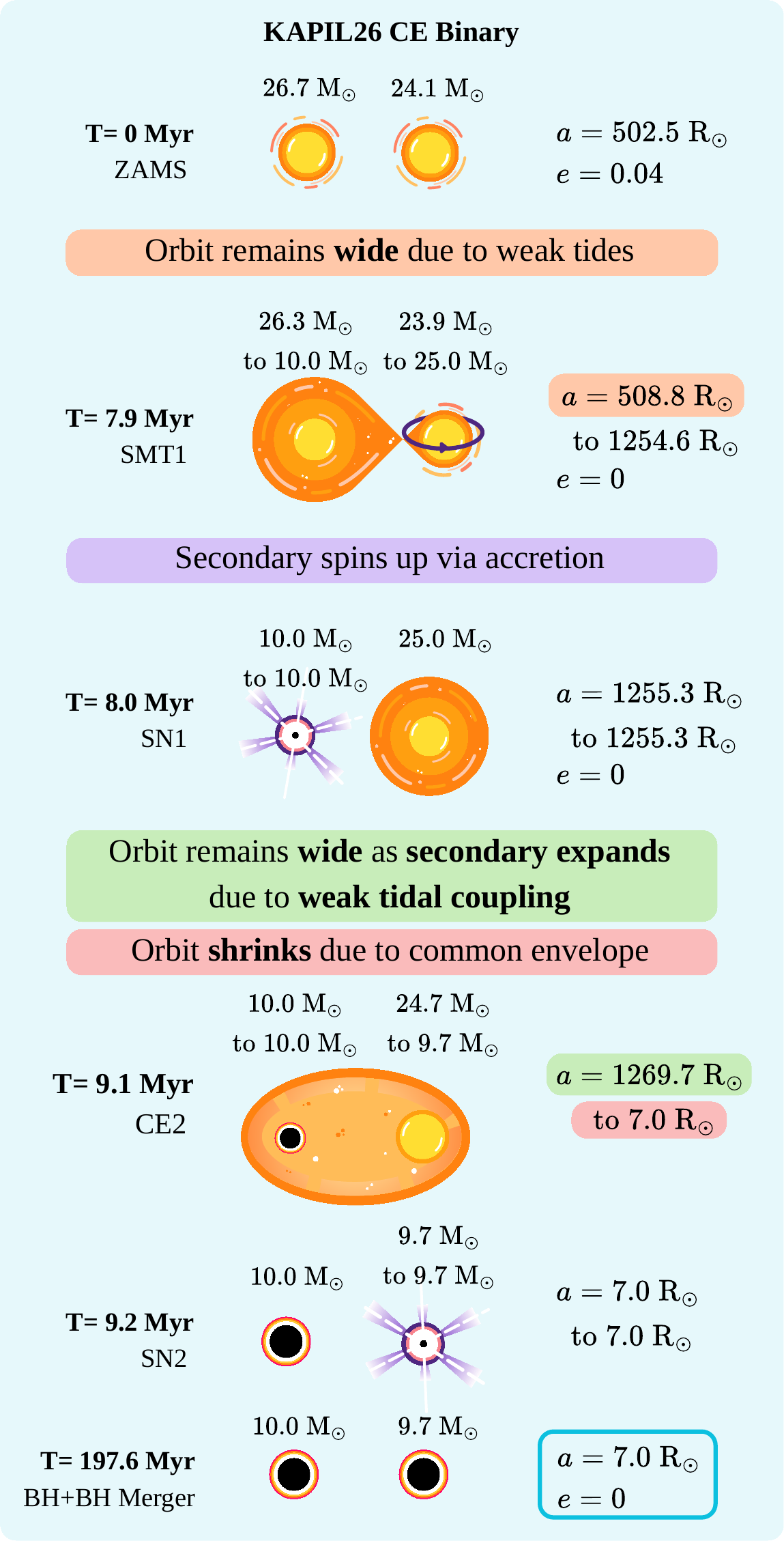}
    \end{subfigure}
    \hfill
    \begin{subfigure}[t]{0.49\textwidth}
        \centering
        \includegraphics[height=6in]{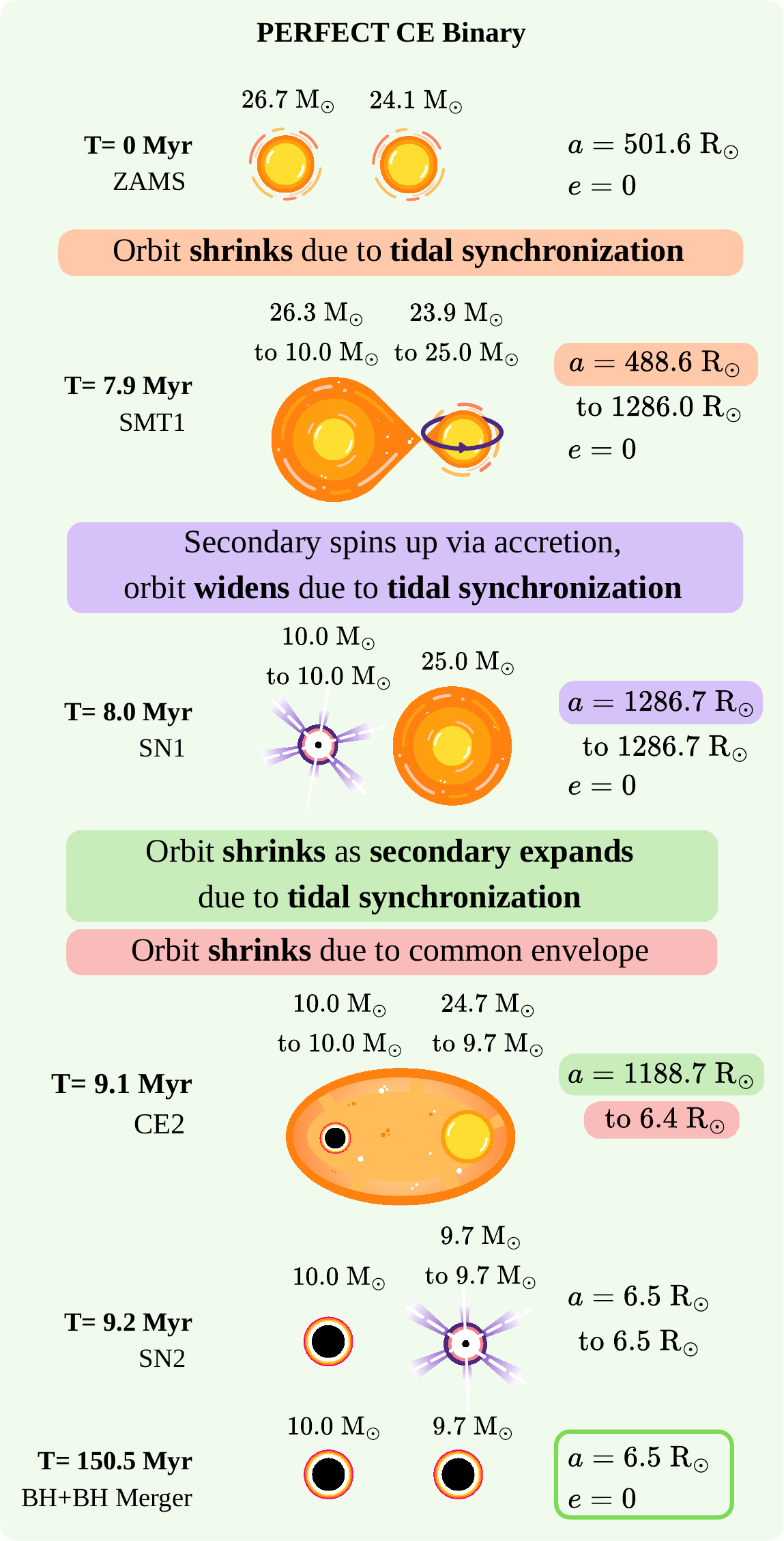}
    \end{subfigure}
    \caption{Evolution of a BBH progenitor that goes through common envelope with both KAPIL26 (left) and PERFECT (right) tides. The initial parameters in both cases are $M_{1, \rm ZAMS} = 26.7 M_\odot$, $M_{2, \rm ZAMS} = 24.1 M_\odot$, $a_{\rm ZAMS} = 502.5 R_\odot$, $e_{\rm ZAMS} = 0.04$, and $Z_{\rm ZAMS} = 0.0002$. The important episodes shown here are: ZAMS, SMT initiated by the primary (SMT1), primary supernova (SN1), common envelope initiated by the secondary (CE2), secondary supernova (SN2), and finally, the BBH merger. Throughout their evolution, neither star develops a significant convective envelope so tides in the KAPIL26 framework operate exclusively through dynamical IGW dissipation and remain weak. In contrast, PERFECT tides instantaneously synchronize the secondary's spin with the orbit, continuously extracting angular momentum from the orbit and transferring it into the star. The result is a wider separation before and after CE2 in the KAPIL26 binary compared to the PERFECT binary, and a longer resulting BBH merger time. This also lowers the BBH merger rate for KAPIL26 tides compared to PERFECT tides.} 
    \label{fig:ce_binary_evolution}
\end{figure*}

The trend of merger yields with tidal prescriptions in Fig.~\ref{fig:branching_ratios_mergers} is rather straightforward. Stronger tides are more efficient at shrinking the pre-CE orbits, leading to BBH merger yields that monotonically increase with tidal strength. To show an example of this behavior, we compare the evolution of two CE binaries with identical initial conditions between the KAPIL26 and PERFECT tidal prescriptions in Fig.~\ref{fig:ce_binary_evolution}. The evolution with the LEGACY prescription proceeds nearly identically to the KAPIL26 model, for reasons that should become clear presently.

Although the initial binary parameters are identical, the PERFECT tides binary is immediately circularized and synchronized in COMPAS, yielding a different effective initial semi-major axis and eccentricity at ZAMS. Both stars in this simulation have convective cores and radiative envelopes at ZAMS. As the primary evolves through the HG and core helium burning (CHeB) phases, a small convective envelope forms outside the radiative shell. As shown in the third row of Fig.~\ref{fig:ce_binary_evolution}, the first phase of stable mass transfer widens the binary more for PERFECT tides than for KAPIL26 tides. This is because the accretor, which is spun up by mass transfer, is able to immediately feed that angular momentum back to the orbit with PERFECT tides. On the other hand, the equilibrium as well as dynamical tides experienced by the star are insignificant according to the KAPIL26 model, and the binary remains tidally decoupled. The only significant tidal contribution comes from  inertial gravity wave (IGW)  dissipation at the convective core – radiative shell boundary, which is still suppressed due to the small core radius. The upshot is that binary separation following the first mass transfer episode is slightly lower with KAPIL26 tides compared to PERFECT tides, caused by absence of tidal coupling with KAPIL26 up to this point.

In both tidal models a CE phase occurs during the second episode of mass transfer (hereby, CE2), initiated by the secondary star. However, the pre-CE2 orbit is appreciably wider in the KAPIL26 model than in the PERFECT model. Prior to CE2, the secondary star evolves past MS and HG, moving onto the CHeB phase. During CHeB, the convective core radius of the primary becomes very small and the radiative envelope radius expands by almost 10 times. In the KAPIL26 tidal model, only dynamical tides due to IGW dissipation act on such a star, since the convective region is too small to experience equilibrium tidal dissipation. These tides are relatively weak due to the small size of the convective core, leading to a very slowly rotating secondary at the onset of CE. The LEGACY model performs very similarly to KAPIL26 tides for such binaries, where tides remain insignificant throughout binary evolution. On the other hand, PERFECT tides, which are agnostic to stellar structure, can efficiently spin up the expanding secondary star and shrink the orbit dramatically prior to CE. This smaller separation carries through the CE phase, all the way to BBH formation. Over our simulated populations, PERFECT tides systematically produce the tightest orbits pre- and post-CE, leading to the highest BBH formation and merger rates.

\subsection{Stable mass transfer}
\label{sec:smt_mergers}
The next significant contribution to BBH formation comes from the channel where the binary experiences only SMT. This channel is much less efficient compared to the CE channel; of all the SMT BBHs formed in our simulations, only $\sim 18\%$ merge within 13.8 Gyr. 
\begin{figure*}
    \centering
    \begin{subfigure}[t]{0.49\textwidth}
        \centering
        \includegraphics[height=6in]{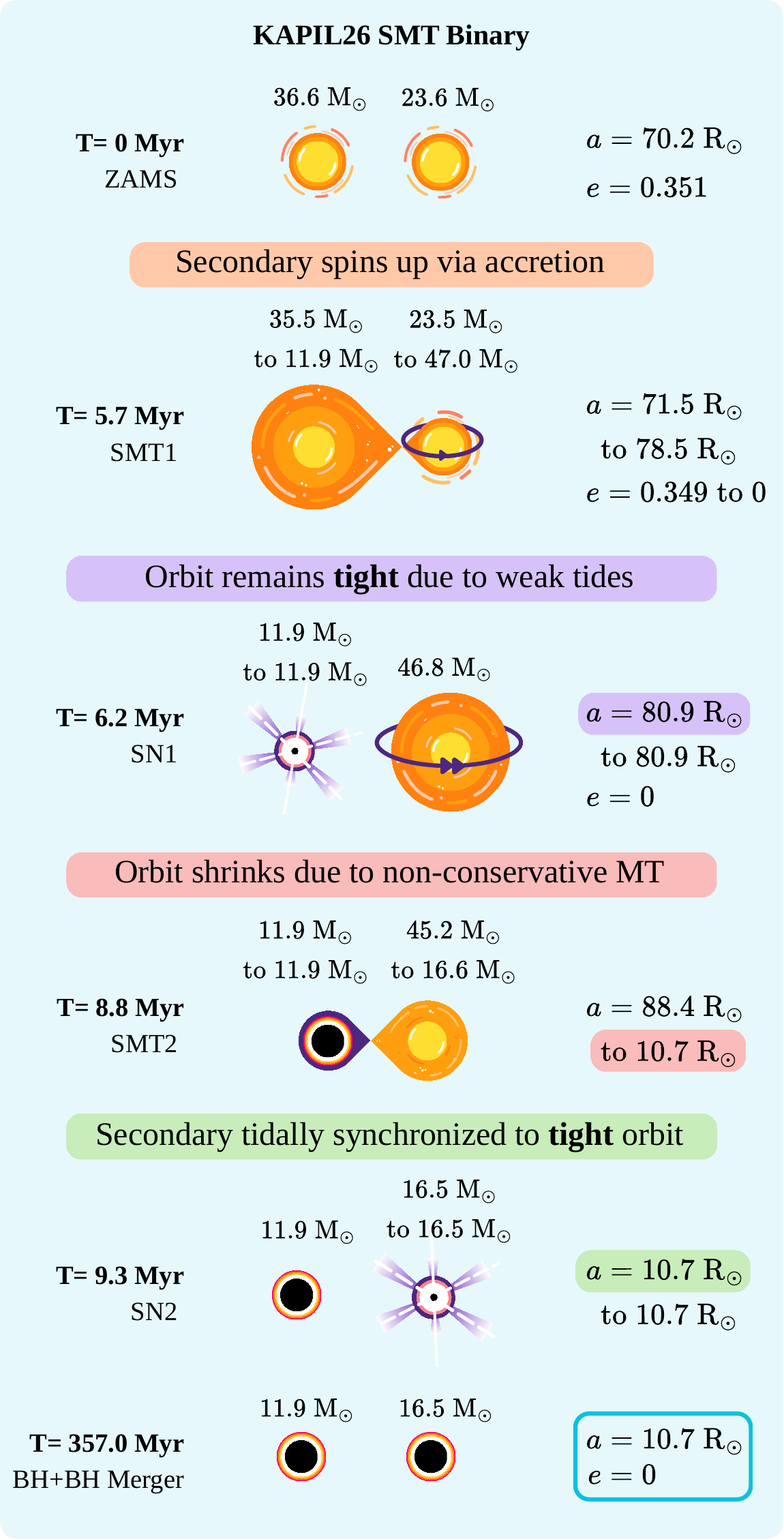}
    \end{subfigure}
    \hfill
    \begin{subfigure}[t]{0.49\textwidth}
        \centering
        \includegraphics[height=6in]{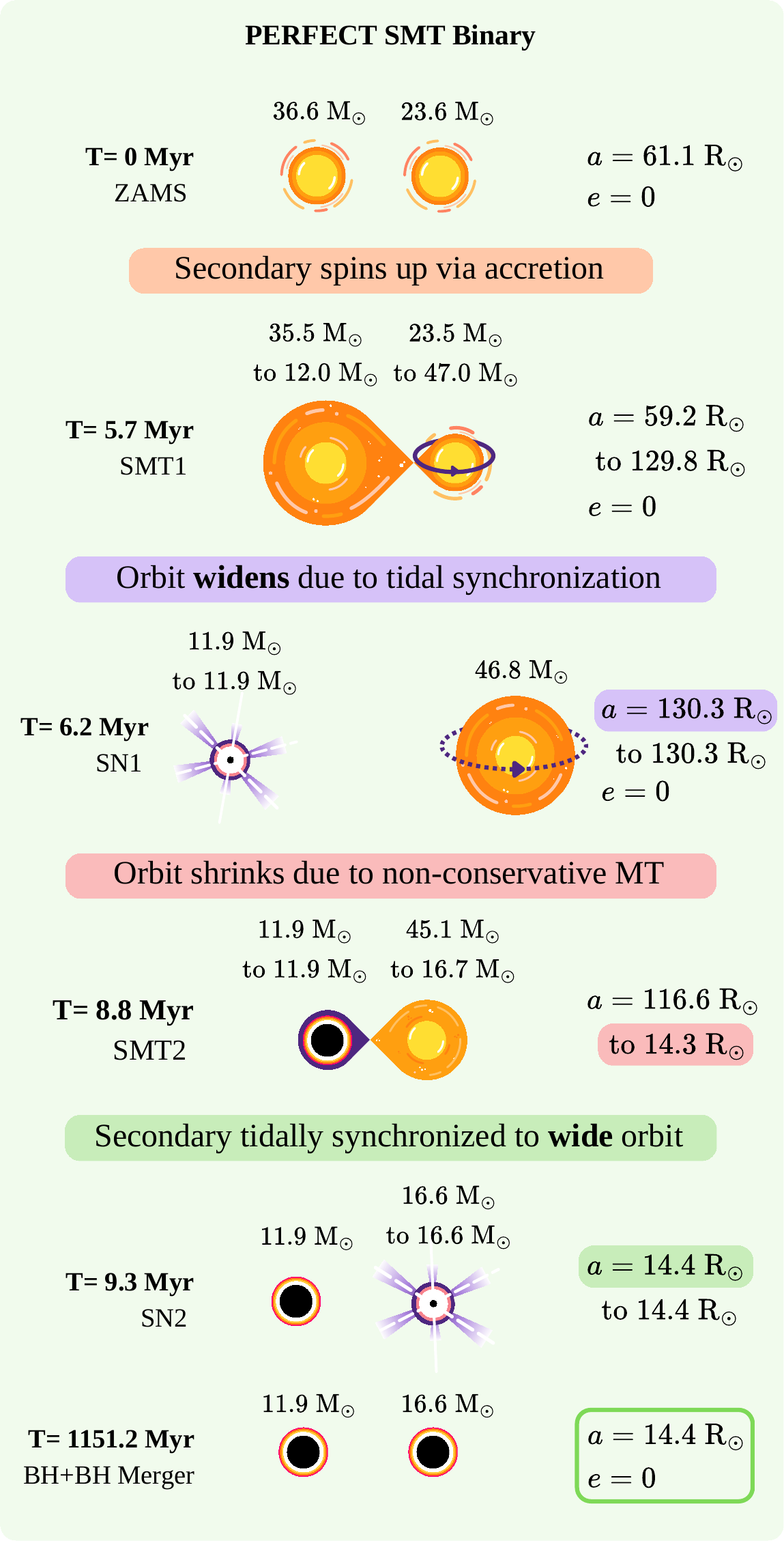}
    \end{subfigure}
    \caption{Evolution of a BBH progenitor that goes through only SMT with both KAPIL26 (left) and PERFECT (right) tides. The initial parameters in both cases are $M_{1, \rm ZAMS} = 36.6 M_\odot$, $M_{2, \rm ZAMS} = 23.6 M_\odot$, $a_{\rm ZAMS} = 70.2 R_\odot$, $e_{\rm ZAMS} = 0.351$, and $Z_{\rm ZAMS} = 0.0005$. The important episodes shown here are ZAMS, SMT initiated by the primary (SMT1), primary supernova (SN1), SMT initiated by the secondary (SMT2), secondary supernova (SN2), and finally, the BBH merger. Throughout the evolution, neither star develops a significant convective envelope, so KAPIL26 tides operate exclusively through dynamical IGW dissipation. The angular momentum deposited into the secondary during SMT1 is not efficiently returned to the orbit by IGW tides, yielding a more compact post-SMT1 separation ($78.5\,R_\odot$) compared to the PERFECT case ($129.8\,R_\odot$) where instantaneous tidal synchronization immediately transfers the excess spin angular momentum back to the orbit, widening the binary and increasing the time to merger. This also suppresses the merger rate.
}
    \label{fig:smt_binary_evolution}
\end{figure*}
During the first SMT event from the primary to the secondary, the accretor gains angular momentum from the in-falling material and may spin up significantly. If mass transfer spins up the accretor beyond orbital synchronization, the eventual outcome for merging BBHs is strongly sensitive to the efficiency with which tides are able to transfer this angular momentum to the orbit and widen it.  Meanwhile, as the moment of inertia of a star increases due to stellar evolution between mass transfer episodes, efficient tides can transfer angular momentum from the orbit to the star, leading to orbital hardening.  This complex interplay, in which tides can cause the orbit to either harden or widen, leads to a complex, non-monotonic relationship between the yields for LEGACY, KAPIL26, and PERFECT tidal models.

These processes are best visualized in Fig.~\ref{fig:smt_binary_evolution}, where we show a schematic evolution diagram of a BBH formed by SMT only with the KAPIL26 and PERFECT models. As in the CE example, stars in this simulation also start with a convective core and a radiative envelope at ZAMS. 

The secondary is spun up during the first stable mass transfer episode (SMT1) in which mass is transferred from the more massive primary to the secondary star. The efficiency of the ensuing tidal dissipation with KAPIL26 tides depends on the secondary stellar structure. Although the secondary develops a thin convective envelope during the HG phase, the mass of the convective envelope is too low for any equilibrium or dynamical tidal dissipation to occur due to the convective envelope as per the KAPIL26 model. Dynamical tides due to IGW dissipation from the convective core - radiative envelope boundary, while non-zero, are also very weak due to the core size. As a result, KAPIL26 tides are too inefficient to return this excess spin angular momentum to the orbit, and the binary retains a relatively compact separation of $78.5~R_\odot$ after mass transfer concludes. Note that the binary also becomes circular immediately following the mass transfer episode, as per default COMPAS assumptions. 

In contrast, the PERFECT tides model instantaneously synchronizes stellar spins with the orbit, so that the angular momentum deposited in the secondary during SMT1 is immediately fed back into the orbit. This is evident in the right panel of Fig.~\ref{fig:smt_binary_evolution}, where PERFECT tides widen the post-SMT1 separation to $129.8~R_\odot$. This wider separation carries through the rest of the binary evolution, leading to wider eventual BBH orbits and lower merger rates with PERFECT tides compared to KAPIL26. Although not shown explicitly, the behavior of LEGACY binaries with similar initial conditions looks almost exactly like the KAPIL26 simulations, where tides do not affect the binary separation.

There is a symmetric set of circumstances that lead to BBHs that form efficiently with KAPIL26 tides and PERFECT tides, but not with the LEGACY model. This effect occurs for binaries that begin their evolution at much smaller orbital separations. At low enough separations, dynamical IGW dissipation in the KAPIL26 model can become efficient, approaching the behavior of the PERFECT model rather than the LEGACY model. In such binaries, the secondary is spun down and the binary is widened efficiently following SMT1 by both the KAPIL26 and PERFECT tidal models. This increase in post-SMT1 orbital separation is often crucial to avoiding a premature stellar merger. Therefore, for small enough initial separations, the LEGACY model produces fewer BBHs via the SMT channel than either the KAPIL26 or the PERFECT models, since several binaries end their evolution as stellar mergers before a BBH can form.

In summary, the non-monotonic SMT BBH yield in Fig.~\ref{fig:branching_ratios_mergers} is best understood as a combination of competing effects at small and large initial separations. For small initial separations, only KAPIL26 and PERFECT tides are able to widen the stellar orbits enough following SMT1 to avoid stellar mergers. For wide initial separations, all three tidal models may produce BBHs but the PERFECT model leads to wider BBH separations, and thus, lower merger rates than the LEGACY and the KAPIL26 models. The KAPIL26 model happens to behave optimally in both situations: widening stellar binaries enough to prevent stellar mergers, and keeping binaries from widening significantly following SMT1 to ensure efficient BBH mergers.

\subsection{Chemically homogeneous evolution}
The last major channel considered in this work is chemically homogeneous evolution (CHE), where both stars in the binary are born close enough and massive enough at ZAMS that they are immediately tidally locked~\citep{deMink:2010zm, Mandel:2015qlu, Marchant:2016, Riley:2020btf}. The resulting high rotation rates drive efficient internal chemical mixing, preventing the development of a hydrogen-rich envelope. These chemically homogeneous stars do not expand and can avoid mass transfer entirely to form compact, rapidly rotating BBHs in situ. The CHE channel is thus important for producing high-mass, high-spin BBHs. Since CHE systems are intrinsically compact, roughly $96\%$ of all CHE BBHs that form also subsequently merge within 13.8 Gyr.

The LEGACY tidal prescription treats CHE stars as being instantly synchronized and circularized, making it equivalent to the PERFECT model for specifically these binaries. Both models therefore predict identical BBH merger rates from the CHE channel. The KAPIL26 model, by contrast, predicts a slightly lower BBH formation and merger rate for CHE binaries. The KAPIL26 model does not make special concessions for CHE stars, evolving them self-consistently in accordance with their stellar structure and binary separation. If any CHE star loses sufficient mass and angular momentum to stellar winds, dynamical tides in the KAPIL26 model may become too weak to support synchronization, and the star may no longer be considered chemically homogeneous. At this point, it would instead resume evolution as an ordinary main sequence star, and roughly $33\%$ of CHE binary stars in the KAPIL26 simulations expand and undergo stellar mergers instead of becoming naked helium stars at the end of the main sequence as in LEGACY or PERFECT.

\section{Impact of Tides on Spin magnitudes}
\label{sec:spin_magnitudes}
A major benefit of implementing tidal spin-up throughout binary evolution and not just in post-processing is that we can obtain the angular momenta and spins of stars at any point in their evolution. For BBH progenitors, we can record the angular momenta of stripped stars right before SN and then map these to the spins of their compact remnants.

There are several uncertainties that complicate DCO spin magnitude predictions. For instance, the rotational velocity of a pre-SN stellar core varies dramatically depending on the assumptions on angular momentum transport between the core and the envelope, which remains uncertain for massive stars \citep{Spruit:1999cc, Heger:1999ax, Qin:2018vaa, Fuller:2019sxi, Ma:2023nrf}. 
In COMPAS, stars are treated as rigid rotators, such that the core and shell spin at the same angular frequency. This is equivalent to assuming highly efficient AM transport within the massive stars which synchronizes the core and the envelope, which may be consistent with theoretical predictions \citep{Fuller:2019sxi, Fuller:2019ckz}. If AM transport is not perfectly efficient, the core will typically rotate faster than the surface and our assumptions would under-predict the spin magnitudes. However, this caveat only affects the BHs that form without complete envelope fallback. We define the fallback fraction as 
\begin{equation}
    f_{\rm fb} = \frac{M_{\rm remnant} - M_{\rm core}}{M_{\rm *} - M_{\rm core}}\,.
    \label{eq:fallback_fraction}
\end{equation} 

Roughly $\sim 85\%$ of the BHs that merge within 13.8 Gyr are formed with $f_{\rm fb}\approx 1$ in our simulations.  While this may suggest that assumptions about angular momentum loss due to incomplete fallback will affect only a relatively small fraction of the merging BBH population, some caution is warranted.  Progenitor stars rotating as rigid bodies will typically have more specific angular momentum in their outer layers than the specific angular momentum at the innermost stable circular orbit around the forming black hole.  Unless angular momentum can be efficiently redistributed during stellar collapse, as we assume here, this could lead to the formation of rotationally supported structures, shocks, and significant additional mass and/or angular momentum loss \citep[e.g.,][]{Murguia:2020}.  Moreover, if the progenitor's angular momentum is larger than the maximum dimensionless value of 1 allowed for a Kerr black hole, some angular momentum must be lost, e.g., by launching jets in a long gamma-ray burst.  We assume that this process is sufficiently finely self-regulated to leave behind a maximally spinning black hole, but this assumption is non-trivial, and more than the minimum necessary amount of angular momentum may be lost.

For the $15\%$ of BHs that experience some mass loss during SN, we must estimate the resulting loss of angular momentum. The merging BHs in our simulations arise from stripped stars that have already lost their hydrogen envelope.  We consider separately the progenitor angular momentum in the core (in practice, we only evolve stars to carbon-oxygen core formation) and the helium shell:
\begin{align}
    J_{\rm core} &= I_{\rm core} \Omega\,, \\
    J_{\rm shell} &= I_{\rm shell} \Omega\,.
\end{align}
Here, $\Omega$ is the rotational velocity of the star, which is evolved self-consistently throughout the star's lifetime in both the KAPIL26 and PERFECT tidal prescriptions. The total angular momentum of the pre-SN star is $J_* = J_{\rm core} + J_{\rm shell}$.  For the LEGACY prescription, $\Omega$ is meaningless since this model does not account for stellar rotation through binary evolution. To remain consistent with other post-processing studies \citep{Zaldarriaga:2017qkw, Bavera:2020inc}, we assume that all stripped stars in the LEGACY simulation are synchronized to the orbital frequency at SN, such that $\Omega = \omega_{\rm orb}$ for each He burning star at the moment of SN. 

The quantities $I_{\rm core}$ and $I_{\rm shell}$ are the moments of inertia of the different regions of the star.  We use the fitting formulae of \citet{Hurley:2000pk}, which are broadly consistent with the more detailed models of \citet{Bavera:2020inc,Marchant:2023ncp,PopaDeMink:2025}:
\begin{align}
    I_{\rm core} &= 0.21 M_{\rm core} R_{\rm core}^2\,, \label{eq:mom_of_interia_core}\\
    I_{\rm shell} &= 0.1 M_{\rm shell} R_{*}^2\,.
    \label{eq:mom_of_interia_shell}
\end{align}

Having approximated the angular momentum stored in the core and the shell under the rigid-body assumption, we can determine the amount of angular momentum accreted onto the remnant after supernova. A truly realistic treatment would require careful consideration of accretion feedback onto the remnant, introducing a non-trivial mapping of initial mass and spin to remnant spin \citep{Batta:2019clm, Qin:2023tbh, Marchant:2023ncp}. COMPAS also does not account for energy lost through gamma-ray bursts, which can carry away additional angular momentum and prevent BHs from reaching maximal spin at birth. In this work, we adopt the simplest and most optimistic model for estimating remnant spins, assuming complete angular momentum transfer of any mass accreted onto the remnant. 
Under this assumption, the remnant angular momentum is a combination of the core and shell angular momenta retained after supernova, i.e.,
\begin{align}
    J_{\rm BH} = J_{\rm core} + (f_{\rm fb} \times J_{\rm shell})\,,
    \label{eq:angular_mom_post_sn}
\end{align}
where $f_{\rm fb}$ is the fallback fraction defined in Eq.~\eqref{eq:fallback_fraction}.
The default supernova kick prescription in COMPAS is set to the \cite{Mandel:2020qwb} model, which determines the mass lost as well as the direction and magnitude of the kick received by the remnant. 

We then follow \citet{Marchant:2023ncp} to regulate the BH spin based on the pre-SN critical rotation rates of stripped stellar progenitors. With this limit, the maximum remnant spin is given by
\begin{equation}
    a_{\rm max} = 10^{c_0 + c_1 x + c_2 x^2}\,,
\end{equation}
where $x = \log_{10}({M_{\rm BH}}/M_{\odot})$, 
\begin{align}
    c_0 &= 0.13 \sqrt{\log_{10}(0.1 Z_\odot / Z)}\,,\\
    c_1 &= 0.74 - 0.19 \sqrt{\log_{10}(0.1 Z_\odot / Z)}\,, \\
    c_2 &= -0.58 +0.12 \sqrt{\log_{10}(0.1 Z_\odot / Z)}\,.
\end{align}  
It is entirely possible that the $a_{\rm max}$ limit from \citet{Marchant:2023ncp} is too pessimistic (see, e.g., \citet{Popa:2025dpz}). However, the more restrictive limit for our simulations comes from the maximal rotation rate for Kerr BHs $a_{\rm BH} \leq 1$. We assume that progenitors that exceed this limit are able to reduce the angular momentum to a spin magnitude of 1 without affecting the remnant mass.
Applying this limit to our BHs, the updated spin can be written as
\begin{equation}
    a_{\rm BH} = \min \left( \frac{c J_{\rm BH}}{G M_{\rm BH}^2}, a_{\rm max}, 1 \right)\,.
\end{equation}

\begin{figure}
    \centering
    \includegraphics[width=\linewidth, trim={1cm 0cm 0cm 0cm}]{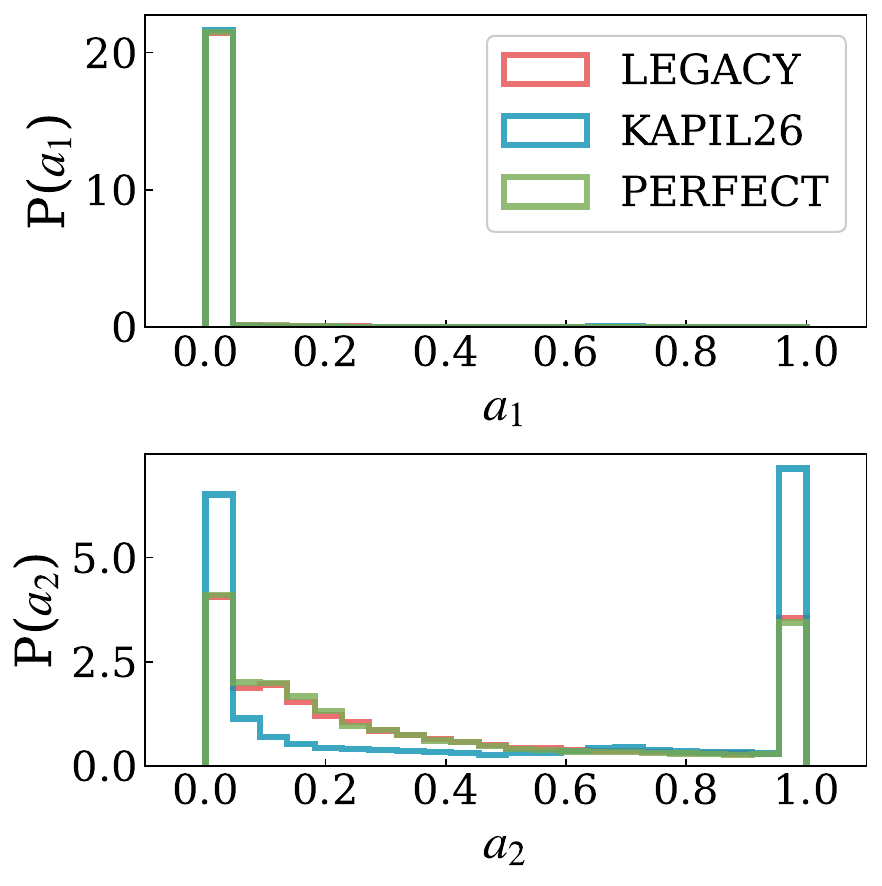}
    \caption{Dimensionless BH spin distributions of the primary (defined as the more massive component at ZAMS) and secondary (defined as the less massive component at ZAMS), across the three simulated populations. The simulations represent a log-uniform initial stellar metallicity distribution, and are not weighted to represent a realistic cosmic star formation history.}
    \label{fig:a1_a2_dist_spin}
\end{figure}

The resulting dimensionless spins for BH remnants in our simulations are shown in Fig.~\ref{fig:a1_a2_dist_spin}. We label the remnants by their ZAMS indices, such that $a_1$ refers to the primary star at ZAMS and not necessarily the more massive BH. This choice allows us to understand our theoretical spin predictions with binary stellar evolution in mind.

In our simulations, the primary star is typically also the first-born BH in the binary, although not necessarily the more massive BH. Unless the BBH is formed via CHE or there was a CE event prior to the first supernova event (SN1), the first BH is born at wide enough orbital separations that tides are unable to rapidly spin up the progenitor star. The resulting BH typically has zero natal spin. On the contrary, the second-born BH can experience a range of tidal and accretion-driven episodes of spin-up prior to the second supernova event (SN2), and can acquire natal spins ranging from 0 to 1, with substantial pile-ups at either end under our simple spin calculations. The peak at zero spin corresponds to a combination of minimal tidal spin-up and negligible fallback during supernova, as the remnant retains little angular momentum from its progenitor. The peak at 1 is due to the Kerr limit, whereby all the binaries with the shortest pre-SN2 orbital periods and largest angular momenta are absorbed into the maximal rotation bin.

\begin{figure}
    \centering
    \includegraphics[width=\linewidth, trim={1cm 0cm 0cm 0cm}]{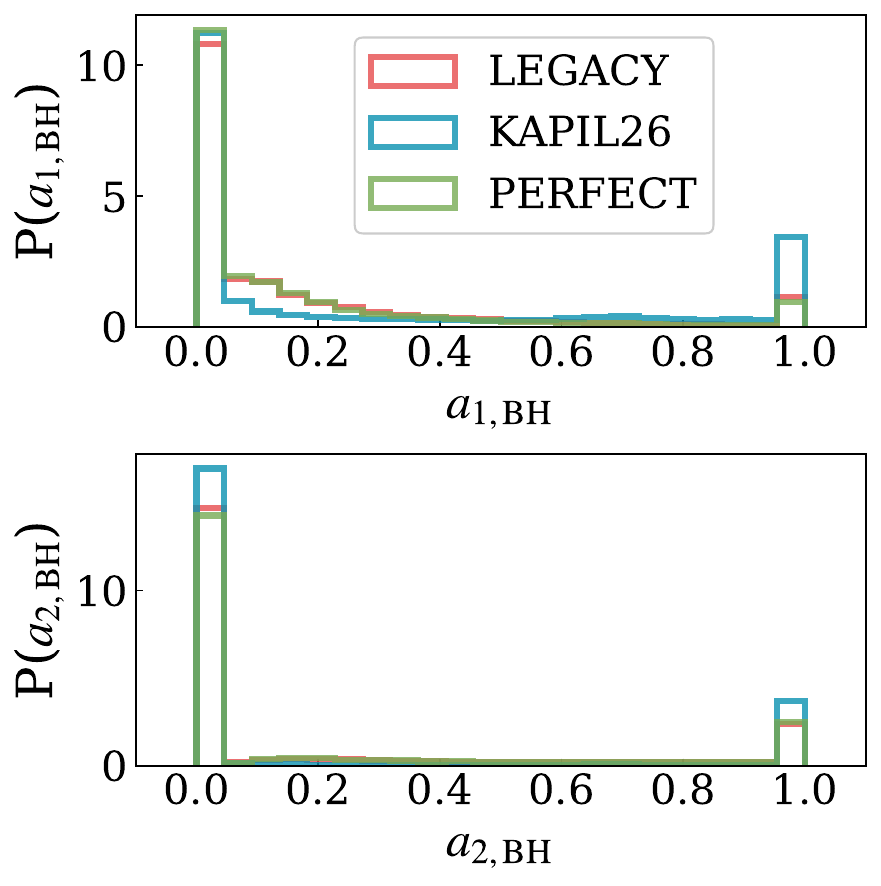}
    \caption{Dimensionless BH spin distributions of the more massive (top panel) and less massive (bottom panel) BH components across the three simulated populations. The simulations represent a log-uniform initial stellar metallicity distribution, and are not weighted to represent a realistic cosmic star formation history.}
    \label{fig:a1_a2_dist_spin_dco}
\end{figure}

Mass ratio reversal (MRR) occurs in roughly $50\%$ of the BBHs that merge within 13.8 Gyr across our simulations, whereby the second-born BH (the erstwhile secondary) becomes the more massive BBH component. Since we only observe compact binaries after stellar evolution has concluded, it is also useful to show the predicted component spin distribution labeled by relative BH mass in Fig.~\ref{fig:a1_a2_dist_spin_dco}. Crucially, we note that either of the observed BBH components can possess non-zero natal spin based on our simulations, which is a result that we see consistently across tidal prescriptions. We will assess the detectable contributions of these rapidly rotating primary BHs in Sec.~\ref{sec:gw_detections}.

\begin{figure}
\begin{subfigure}[t]{0.47\textwidth}
        \centering
        \includegraphics[width=\textwidth]{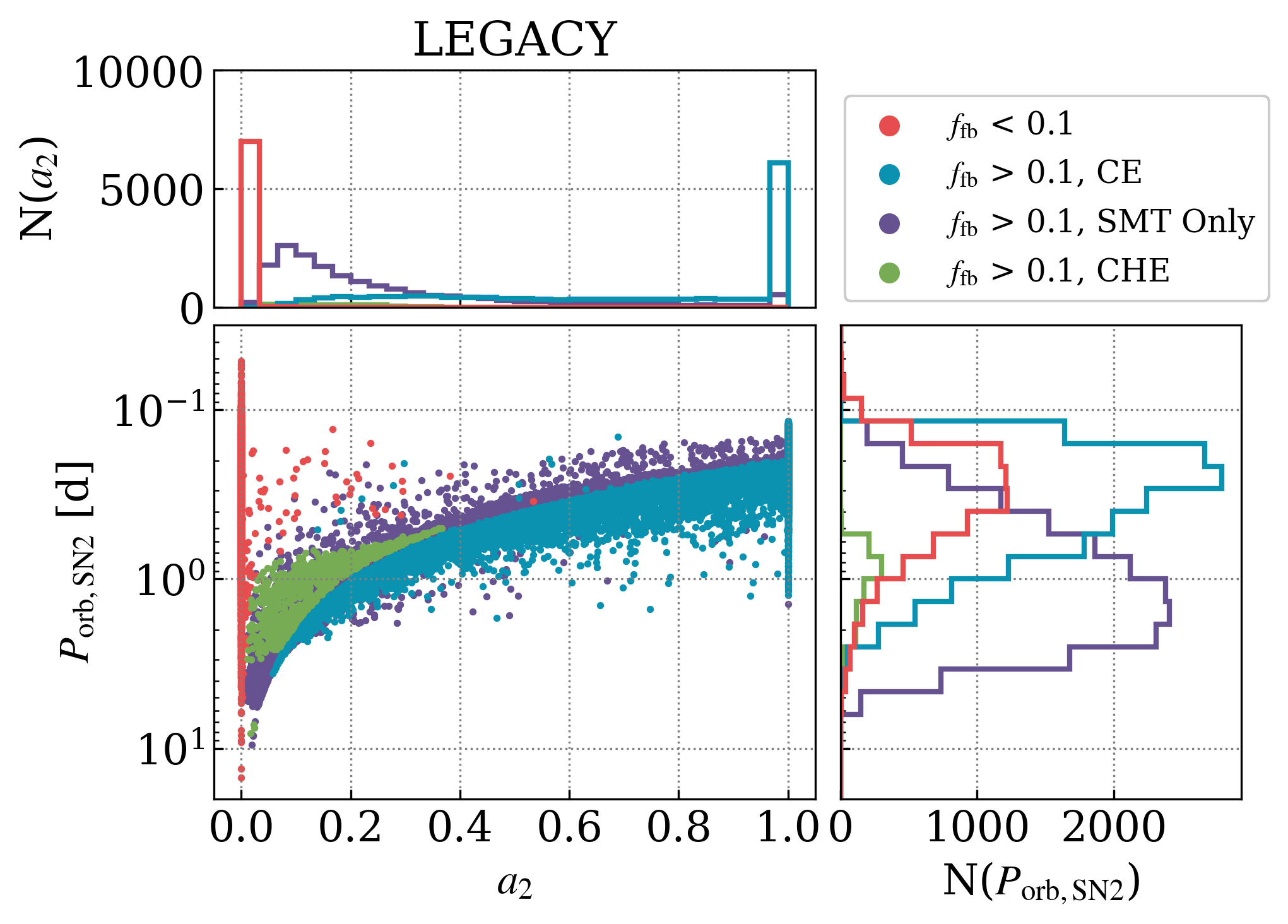}
        \label{}
    \end{subfigure}
    \hfill
    \begin{subfigure}[t]{0.47\textwidth}
        \centering
        \includegraphics[width=\textwidth]{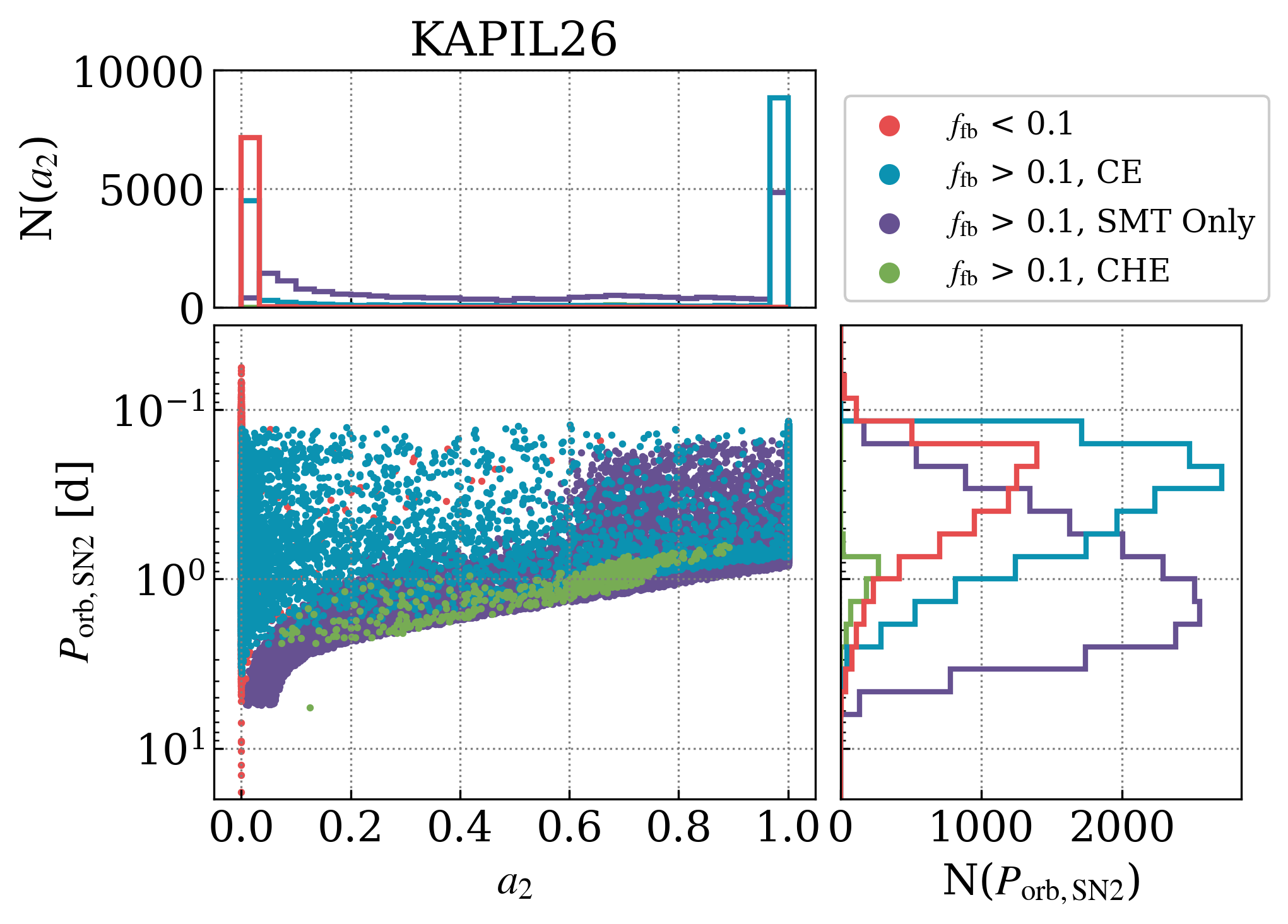}
        \label{}
    \end{subfigure}
    \hfill
    \begin{subfigure}[t]{0.47\textwidth}
        \centering
        \includegraphics[width=\textwidth]{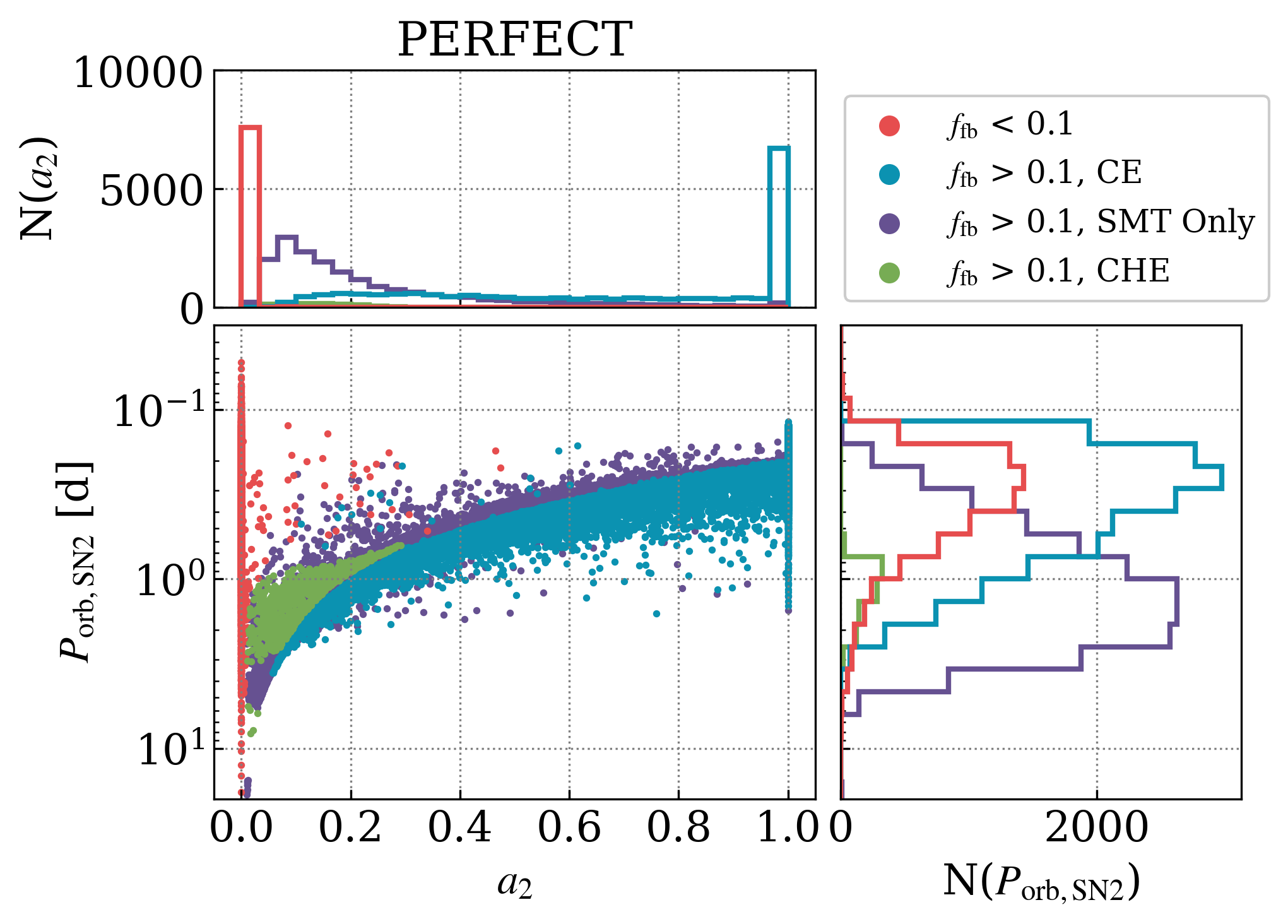}
        \label{}
    \end{subfigure}
    \caption{Dimensionless secondary BH spins (with secondary defined as the less massive component at ZAMS) vs. orbital period prior to secondary SN (note reversed axes), for the BBHs that merge within 13.8 Gyr as obtained with the LEGACY (top), KAPIL26 (middle), and PERFECT (bottom) simulations. The red data points correspond to the secondary receiving fractional fallback $f_{\rm fb} < 0.1$ after SN2, whereas the blue, purple, and green data points all have $f_{\rm fb} > 0.1$ and represent the CE, SMT, and CHE formation histories, respectively.}
    \label{fig:a2_period2_comparison}
\end{figure}

For the BBHs that merge within 13.8 Gyr of formation, we show a breakdown of the spins and pre-SN2 orbital periods by various formation scenarios for our three populations in Fig.~\ref{fig:a2_period2_comparison}. This plot emphasizes the difference between imposing pre-SN2 synchronization and allowing the spins to be determined self-consistently throughout binary evolution. Generally speaking, the spins in the LEGACY and PERFECT models are well correlated to the orbital periods prior to SN2 by construction. However, spins in the KAPIL26 model rarely follow neatly from the pre-SN2 orbital period, instead acquiring a large spread due to the combined effects of mass transfer, stellar evolution, and tidal evolution. 
Once again, we find it useful to categorize the binaries into the CE, SMT, and CHE formation channels. Depending on the specific formation pathway, the spins obtained by assuming pre-SN2 synchronization may be lower or higher than our self-consistent estimates. We will examine these trends more thoroughly in the remainder of this section.

Note that the orbital period axes have been flipped in the central panels of Fig.~\ref{fig:a2_period2_comparison} to preserve the usual orientations of the right-hand panels. When considering only the BBHs that merge within 13.8 Gyr, all three prescriptions have largely similar orbital separations prior to SN2 (see right-hand panels of Fig.~\ref{fig:a2_period2_comparison}). So the differences in the resulting spin distributions come down to the interactions between tides and mass transfer histories.

\subsection{Common envelope}
For binaries undergoing a CE phase, the LEGACY and PERFECT models both predict very high secondary spins, with significant pile-ups at $a_2 \simeq 1$. In addition to the high-spin peak, the KAPIL26 model predicts a much larger number of low-spin secondaries for any given pre-SN2 orbital period.

It is perhaps easiest to understand the trends for the LEGACY and PERFECT tidal models, where the $a_2$ distribution is determined by assuming, or explicitly imposing, synchronization to the orbital period prior to SN2. For these populations, the resulting spin distribution follows a tightly clustered relation with the orbital period, and any scatter arises from the variance in the pre-SN value of $I_2/M_2^2$.

The KAPIL26 prescription results in the largest $a_2$ spread among all the tidal prescriptions considered in this work. Most notably, the KAPIL26 model produces a sizable population of non-spinning secondary BHs from CE evolution, despite the BHs experiencing non-negligible fallback. After envelope stripping during CE, the secondary star typically becomes a stripped helium main sequence (HeMS) star with an entirely radiative envelope and a convective core. If the convective core radius is too small, the minuscule core-envelope interface may lead to drastically lower efficiency of dynamical tidal dissipation via IGWs. In this case, tides can be too weak to synchronize the stellar rotation to the newly increased orbital rate following the CE event, and the secondary rotation rate is not able to increase before SN2. 
There are also binaries where dynamical tides can be much stronger, particularly when the convective core is large enough. Efficient IGW dissipation may lead the spin periods to become (nearly) synchronized with the pre-SN2 orbital periods, contributing to the pile-up at $a_2=1$ with a wide spread.

The above comparisons highlight a key prediction of realistic tidal modeling:  assuming that post-CE stars are tidally synchronized to the orbital frequency can be a strong, if not misleading, imposition. When we account for the impact of stellar structure on dynamical tides, we find that a significant number of second-born post-CE BHs are born with zero spins, even if they experience complete fallback. 

\subsection{Stable mass transfer}
The leading order contribution to the spins of SMT binaries comes from their mass transfer history. As described in Sec.~\ref{sec:smt_mergers}, when the primary star overflows its Roche lobe, the secondary star typically experiences significant accretion of both mass and angular momentum. The efficiency of various tidal prescriptions to transfer this angular momentum from the secondary to the orbit determines how rapidly the secondary will be rotating prior to BH formation.

The LEGACY model does not include any mechanism of angular momentum transfer via tides, and so, the secondary spin is determined largely by the pre-SN2 orbital period. The resulting $a_2$ distribution peaks around $a_2\approx 0.1$, with a tail extending to higher spins. PERFECT tides are able to instantly extract angular momentum from the secondary following accretion, leading to wider orbital separations post-SMT. After all the stellar and binary evolution steps, the differences in the pre-SN2 orbital separations is not vast enough to dramatically impact the $a_2$ distribution. The secondary spin distribution with PERFECT tides remains similar to the LEGACY model, only with slightly lower spins.

The KAPIL26 model produces the highest number of maximally rotating second-born BHs from the SMT channel. This is largely because the KAPIL26 tidal dissipation is often not strong enough to dissipate the excess angular momentum from the secondary star into the orbit during the MS or HG phases. The result is a smaller orbital separation after SMT as well as excess angular momentum in the secondary star. Stellar contraction and some tidal coupling may lead these secondary stars to have high rotation rates at BBH formation, enhancing the $a_2 \simeq 1$ population. In the simulation shown in Fig.~\ref{fig:smt_binary_evolution}, for instance, the secondary retains a large fraction of its excess angular momentum during the MS and HG phases, and therefore remains spinning rapidly during the HeMS and HeHG phases. Tides are unable to widen the post-SMT1 orbit and synchronize the secondary in $\sim 60\%$ of the SMT BBH mergers in our simulations, making this the dominant outcome for SMT binaries.

However, if tidal synchronization (typically due to dynamical tides) is efficient, the secondary can be spun down while on the MS or HG, widening the orbit. Over the course of its subsequent evolution, the secondary may overflow its Roche lobe and initiate a non-conservative SMT episode which lowers the orbital separation. Tides are not always efficient at synchronizing the stripped secondary star to the newly shrunk orbit, and the final spins of these secondary stars can remain low as they were set by the pre-SMT2 orbital rates. These binaries contribute to a wide range of spins for the KAPIL26 simulation, rather than a distinct pile-up at $a_2 \approx 0.1$. 

Once again, depending on the specifics of binary evolution and mass transfer history, the assumption of tidal synchronization of the secondary star prior to core-collapse may be a poor one. In particular, the KAPIL26 tidal model can predict much higher as well as much lower rotational periods for the secondary star compared to the orbital period at core collapse.

\subsection{Chemically homogeneous evolution}
Although the CHE channel may not be very prominent in our simulations, CHE binaries are notable for consistently producing both BHs with non-zero spins. The ZAMS orbital periods of the CHE stars are typically $\mathcal{O}(1)$ day, and CHE stars are tidally locked to their orbital frequencies in all three tidal models, either by self-consistent tidal synchronization (as in KAPIL26) or by explicit assumption (as in PERFECT and LEGACY).

The final spins of the BHs formed from CHE progenitors depend on a few factors. If there is very little fallback following supernova, BHs may lose a significant fraction of their angular momenta, and thus have low natal spins regardless of the tidal model. For non-negligible fallback, both the LEGACY and the PERFECT models produce CHE BBHs with spins that are consistent with their pre-SN orbital periods by construction, as seen in the clustering of the green dots in the top and bottom panels of Fig.~\ref{fig:a2_period2_comparison}. Both these tidal models assume tidal synchronization even after the stars evolve off the CHE phase. However, this may not be a realistic assumption. CHE stars at the end of MS can have significant angular momenta, and although some angular momentum can be lost through stellar winds and tidal spin-down, a large fraction may be retained throughout subsequent evolution. This is precisely what we see in the self-consistent KAPIL26 model, where tidal coupling is often too weak to spin down the evolved (HeMS and HeHG) stars that retain angular momentum from their CHE phase. Therefore, for any given pre-SN2 orbital period, the KAPIL26 model tends to produce higher spin BHs than the other two prescriptions. This effect is negated in the other two prescriptions, where the final stellar rotational period is always imposed to be equal to the orbital period. Once again, we find that the assumption of tidal synchronization in evolved stars can be too strong, and lead to systematic over-estimation of BH spins for CHE binaries.

\section{Astrophysical BBH Spin Distributions}
\label{sec:bbh_astro_population}

\begin{figure}
    \centering
    \includegraphics[width=\linewidth, trim={1cm 0cm 0cm 0cm}]{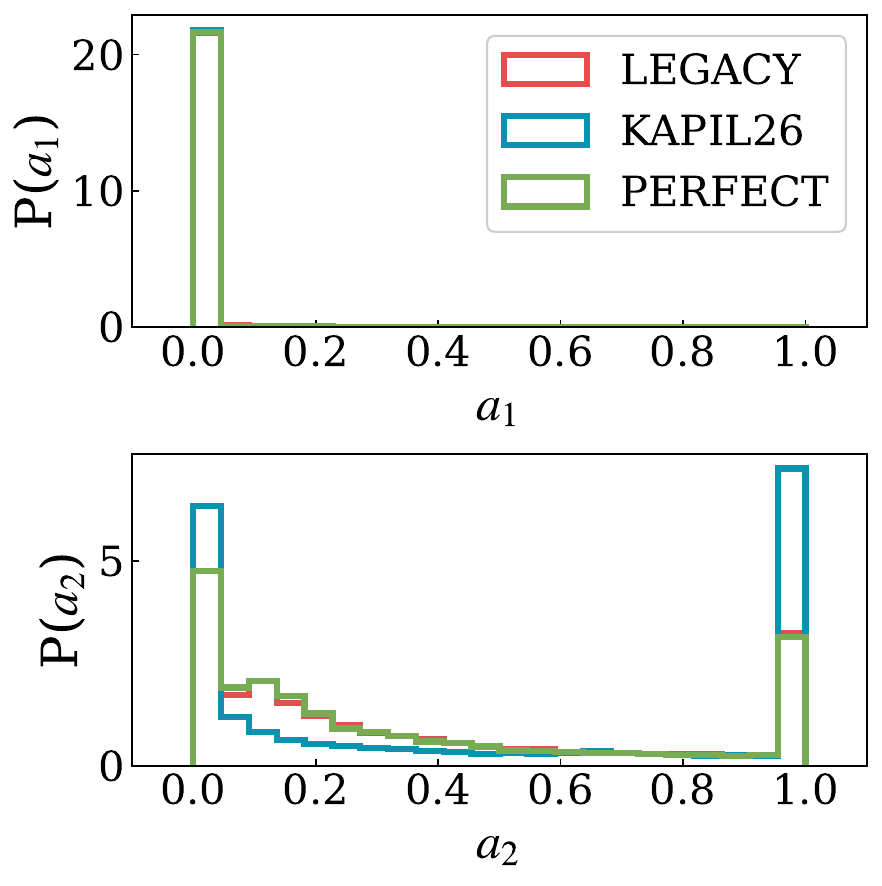}
    \caption{Dimensionless BH spin distributions of the primary (defined as the more massive ZAMS star) (top panel) and secondary (bottom panel) for the merging BBHs across the three tidal models. The spin distributions are weighted by the appropriate metallicity-specific star formation rates.}
    \label{fig:a_bh_population_comparison}
\end{figure}

\begin{figure*}
    \centering
    \includegraphics[width=\linewidth, trim={1cm 0cm 0.5cm 0cm}]{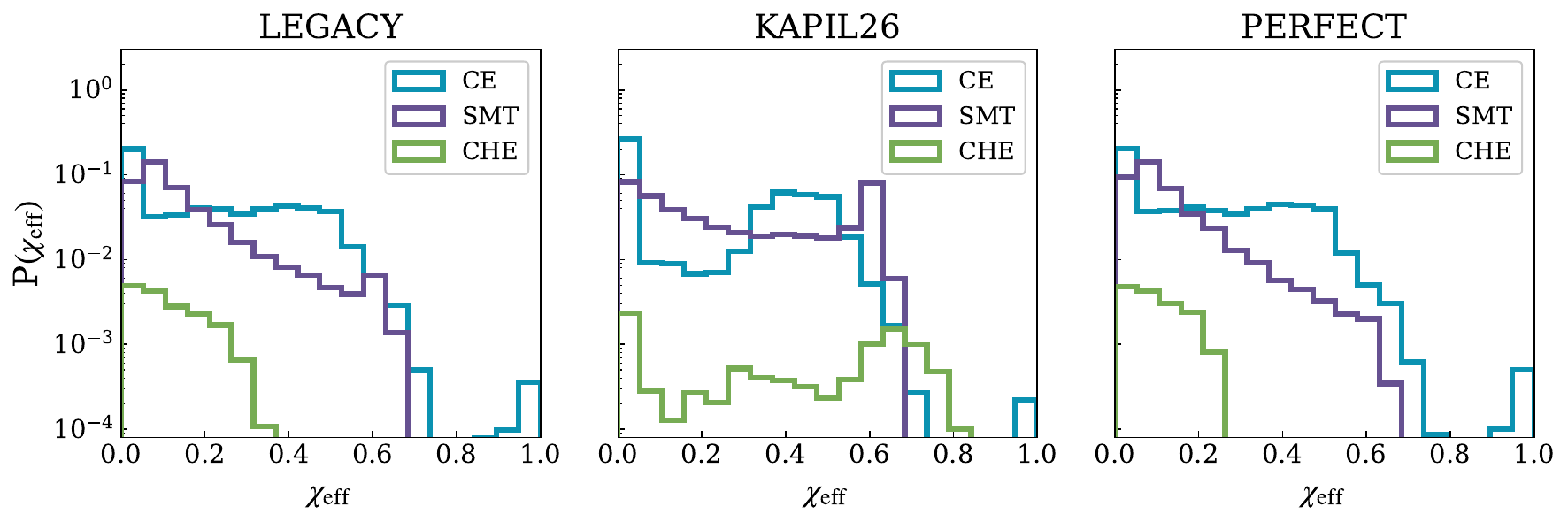}
    \caption{$\chi_{\rm eff}$ probability distributions of BBH mergers with the LEGACY (left), KAPIL26 (center) and PERFECT (right) tides models, broken into sub-populations corresponding to CE, SMT, and CHE formation channels.}
    \label{fig:chi_eff_comparison}
\end{figure*}

In previous sections, we identified and discussed several promising features in BBH spins across our simulations.  However, we have so far considered the properties of all ultimately merging BBHs that would arise from a log-uniform distribution of metallicities.  In reality, although there is a broad metallicity distribution at any formation redshift, the average metallicity rises as the Universe ages.  Higher metallicities that are more prevalent at low formation redshifts lead to higher winds, reducing BH masses and widening BBHs, which in turn may reduce the fraction of rapidly spinning binaries. Some of the channels that yield particularly rapid spins, like CHE, are suppressed at high metallicities, i.e., at low star formation redshifts.  Moreover, as discussed below, rapidly spinning BBHs preferentially merge quickly once formed. Therefore, we should expect the properties of the BBH population to evolve non-trivially as a function of merger redshift.

In order to study the expected distributions of BBH spins over a realistic population, we convolve the outputs of COMPAS simulations introduced in Sec.~\ref{sec:population_details} with the metallicity specific cosmic star formation history model introduced by \citet{Neijssel:2019irh}, following the procedure described in that work and in \citet{COMPAS_2022}. While this approach relies on a simplistic model of the star formation history and metallicity evolution \citep[see, e.g.,][for discussions]{Chruslinska:2019, vanSon:2022}, the overall trends should remain robust.

\subsection{1D Spin Distributions}
\label{sec:1d_spins}

We show the expected intrinsic component spin distributions for the three tidal models considered in this work in Fig.~\ref{fig:a_bh_population_comparison}. 
These spin distributions are obtained by weighting each merging BBH based on its mass and metallicity contribution to the total star forming mass in the Universe, and then integrating over cosmological merger redshifts. In this way, the distributions shown in Fig.~\ref{fig:a_bh_population_comparison} represent the expected intrinsic spin distributions over all merging BBHs that may form in the Universe, although not the detectable distributions. 
MRR still affects about half the merging BBH population, and MRR in the KAPIL26 population ensures that the more massive BH can be born with maximal rotation in $\sim 15\%$ of the merging BBH systems. For comparison, the less massive BH is born with maximal spin in $\sim 30 \%$ of merging BBHs. 

Supernovae can tilt the spin axis with respect to the binary orbital angular momentum~\citep{Kalogera:1999tq, OShaughnessy:2017eks, Baibhav:2024rkn}. This relative inclination is useful to determine the effective spin parameter for the final BBH,
\begin{equation}
    \chi_{\rm eff} = \frac{m_1 a_1 \cos\theta_1 + m_2 a_2 \cos\theta_1}{m_1 + m_2},
\end{equation}
where $\theta_1$ and $\theta_2$ are the angles that the primary and secondary spin vectors make with the orbital angular momentum, respectively. The effective spin  $\chi_{\rm eff}$ is a useful parameter when discussing populations of BBHs, since it is typically measured better with GWs than individual component spins. Our tidal models do not account for inclination angles when determining tidal dissipation strength, but COMPAS computes the inclination angle of a post-SN star relative to the binary orbital angular momentum after each supernova event. The default supernova model in COMPAS determines the kick magnitude following \citet{Mandel:2020qwb}, with the polar and azimuthal kick angles being drawn from an isotropic distribution. In the results going forward, we use these post-SN inclination angles to determine $\chi_{\rm eff}$ for the merging BBHs in each of our simulations. We show the resulting $\chi_{\rm eff}$ distributions from all three tidal simulations, separated into various formation channels, in Fig.~\ref{fig:chi_eff_comparison}. Once again, the histograms are weighted by metallicity-specific star formation history to represent a realistic intrinsic population of BBHs.

Since the second-born BH is usually the only component with non-negligible spin (except for CHE binaries), we can easily understand the $\chi_{\rm eff}$ distribution in Fig.~\ref{fig:chi_eff_comparison} by considering the dominant mass ratios of the BBHs in each formation scenario. For BBHs formed via CE in which the secondary reaches $a_2~\simeq~1$, the BBH mass ratio is approximately in the range $0.7~\lesssim~ q \equiv M_{2}/M_{1}~\lesssim~1.0$, which, under the pairing $(a_1~=~0,\ a_2~=~1)$, places the effective spin peak at $0.4~\lesssim~\chi_{\rm eff}~\lesssim 0.5$. MRR affects $\sim 20 \%$ of the merging CE binaries such that the second-born BH is also the more massive BH. When accompanied by maximal $a_2$, the $\chi_{\rm eff}$ values can exceed 0.5 for these systems. 
There is also an exceedingly rare set of CE BBHs with $\chi_{\rm eff}=1$ which exists across all three tidal simulations. This corresponds to the situation where the CE from the secondary star takes place before the primary star can experience supernova, leading to very tight orbital separations prior to the formation of both BHs.

For SMT only binaries, the $\chi_{\rm eff}$ distribution of merging BBHs across all three prescriptions is dominated by systems that have undergone MRR, yielding $q \equiv M_2/M_1 > 1$. Under the pairing mode of $(a_1~=~0, a_2~=~1)$, a mass ratio of $q=1.5$ places the high-spin peak at $\chi_{\rm eff} \simeq 0.6$ for KAPIL26 and LEGACY tides.

The CHE channel often produces BBHs in the $\chi_{\rm eff} > 0.5$ region in the KAPIL26 population due to non-zero spins of both BHs. However, the astrophysical weighting procedure reveals a very low contribution of CHE binaries to the overall population of merging BBHs across tidal prescriptions.

\subsection{Correlations across binary parameters}
\label{sec:correlations}

\begin{figure*}
    \centering
    \includegraphics[width=\linewidth]{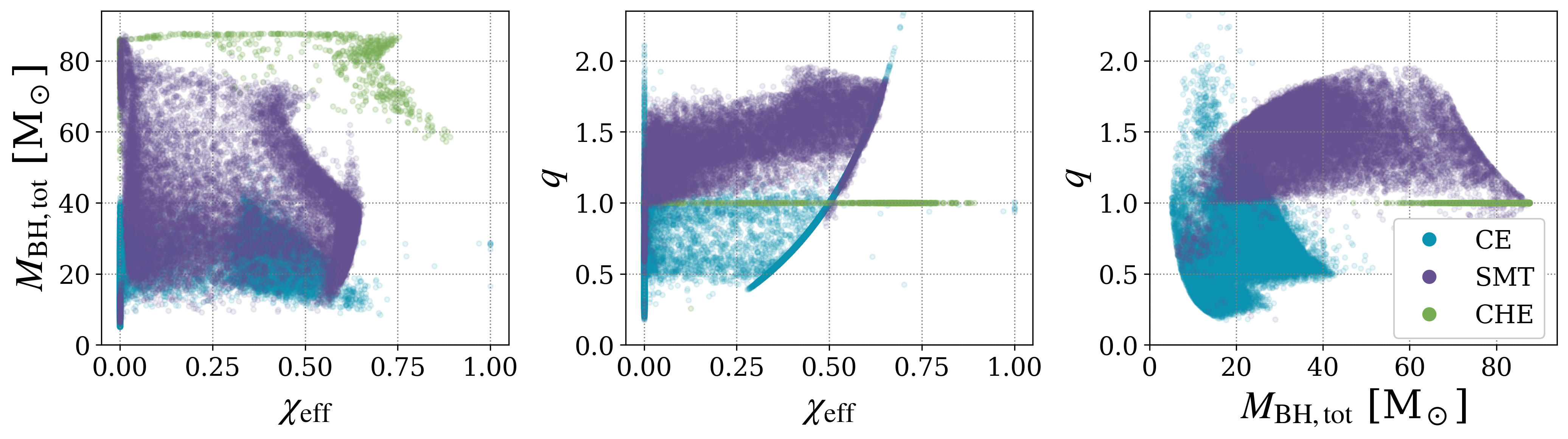}
    \caption{For the KAPIL26 population only: total BBH mass ($M_{\rm BH,tot} = M_{\rm 1} + M_{\rm 2}$) vs. mass ratio ($q = M_{\rm 2} / M_{\rm 1}$) vs. effective spin ($\chi_{\rm eff}$) for BBHs. $q>1$ indicates that the binary experienced mass ratio reversal. All the BBHs merge within 13.8 Gyr of ZAMS formation.}
    \label{fig:3d_subpopulations}
\end{figure*}

\begin{figure}
    \centering
    \includegraphics[width=0.9\linewidth, trim={35 0 0 0}]{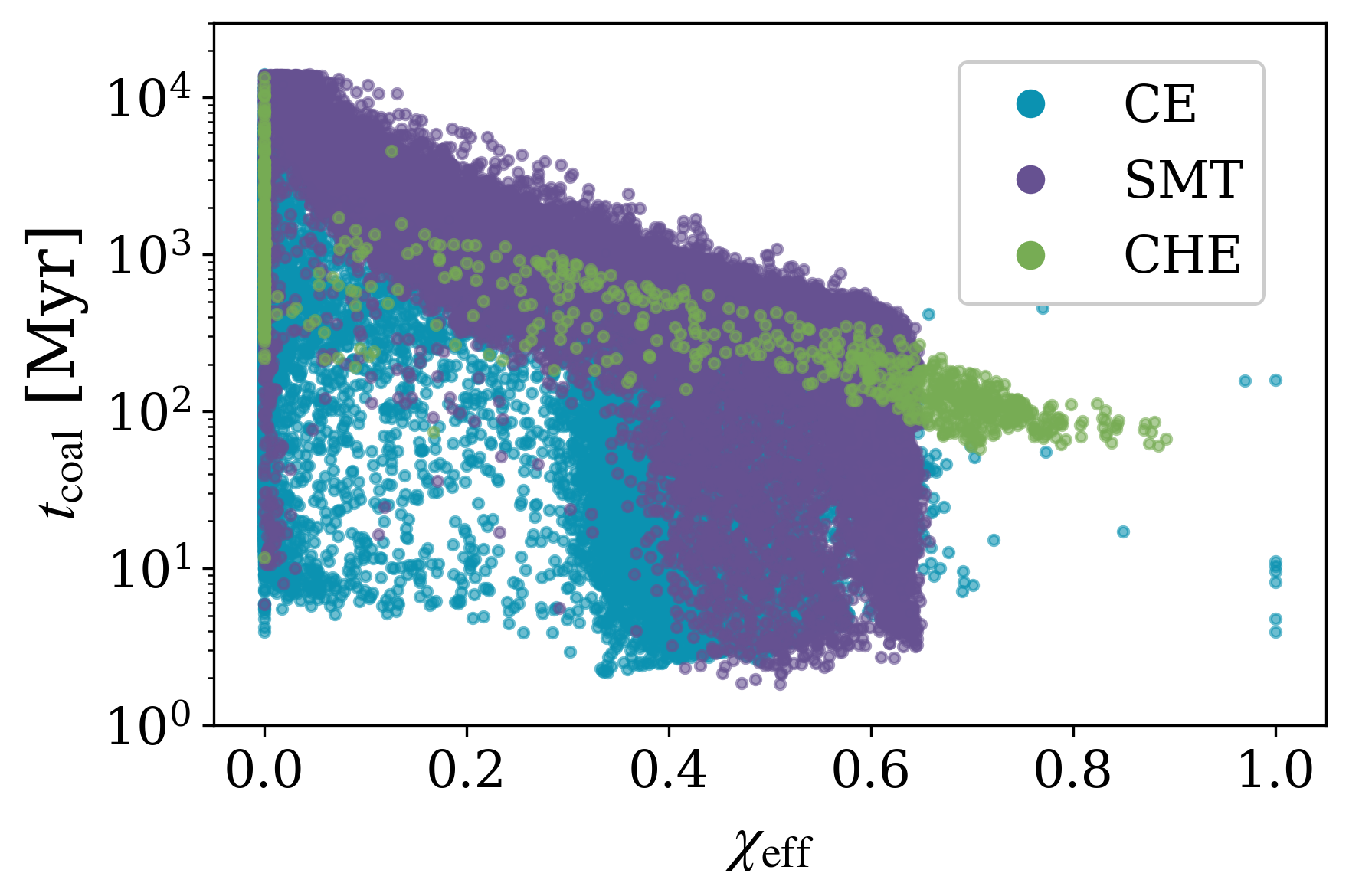}
    \caption{For the KAPIL26 population only: effective spin ($\chi_{\rm eff}$) vs. coalescence time ($t_{\rm coal}$) for all BBHs that merge within 13.8 Gyr.}
    \label{fig:tcoal_vs_chi_eff}
\end{figure}

For the KAPIL26 population only, we show a joint scatter plot of of $\chi_{\rm eff}$, $M_{\rm tot} = M_{1} + M_{2}$, and $q = M_{2}/M_{1}$ in Fig.~\ref{fig:3d_subpopulations}. Note that we have defined $q$ as the ratio of the secondary star's remnant mass to the primary star's remnant mass, which is not directly measurable. However, this choice makes it easy to identify systems that experienced mass ratio reversal as having $q>1$, i.e., where the secondary star became the more massive BH. This figure serves to illustrate how different formation scenarios give rise to BBHs with specific parameter configurations, at least for the KAPIL26 tidal model. 

As we can observe in Fig.~\ref{fig:3d_subpopulations}, the CE binaries are confined to $M_{\rm tot} \lesssim 40\,M_\odot$. These binaries also extend quite far in $q$, reaching below $q=0.5$ in some cases. This may complicate the identification of isolated BBH formation histories based on mass ratio and spins, as has been suggested previously \citep{Banerjee:2024wbq, Olejak:2024qxr}.

SMT binaries occupy a distinct section in $M_{\rm tot}~-~q~-~\chi_{\rm eff}$ space. Not only can SMT BBHs be found with $M_{\rm tot}~\gtrsim~40 M_\odot$, above CE BBHs, but they also dominate the $\chi_{\rm eff}\gtrsim 0.5$ region due to the contribution of MRR binaries. 
For $a_1 \approx 0$ and $a_2 \approx 1$, the effective spin parameter can be as large as $\chi_{\rm eff} \approx q/(1+q)$. This leads to a naturally correlated population of SMT BBHs in the $q$-$\chi_{\rm eff}$ plane, in agreement with previous results from isolated binary evolution \citep{vanSon:2021zpk, Banerjee:2024wbq,Olejak:2024qxr}. 

CHE systems have distinctly equal mass ratios ($q~\approx~1$), and also $M_{\rm tot} \gtrsim 60 M_\odot$. Combined with their high spins under the KAPIL26 model, these properties make them relatively easy to identify from other isolated evolutionary scenarios.

In agreement with previous studies, binaries with large $\chi_{\rm eff}$ preferentially have small coalescence times between BBH formation and merger, since they are born at smaller orbital separations \citep{Kushnir:2016zee, Zaldarriaga:2017qkw, Qin:2018vaa, Bavera:2020inc}. We show this explicitly in Fig.~\ref{fig:tcoal_vs_chi_eff}.  This is not a feature unique to any particular channel, but rather a fundamental consequence of the underlying binary physics: systems that are tidally spun up to high $\chi_{\rm eff}$ are precisely those that merge rapidly. As we will show in the next section, many of these systems will merge at high redshifts, placing them at cosmological distances beyond the reach of current GW detectors. This is precisely the behavior seen in Fig.~\ref{fig:chi_eff_stacked}, which shows the distributions of $\chi_{\rm eff}$ for binaries at several merger redshifts: $z=0, 1, 2, 5$.

\subsection{Uncertainties with tides}

Our models incorporate prescriptions for tidal dissipation via equilibrium tides damped by convection and gravity waves excited at a convective core boundary as described in \cite{Kapil:2026hiw}. These prescriptions could underestimate tidal dissipation for HG stars where there is no convective core or envelope, but there could be a convective H-burning shell. This would allow more rapidly rotating secondary star accretors to be tidally spun down, reducing the number of $a_2 \simeq 1$ BHs. During the CHeB phase, our prescriptions may overestimate tidal dissipation, as \cite{Ma:2023nrf} found that gravity waves often form standing modes, rather than the efficiently damped traveling waves assumed by our prescription. This could prevent slowly rotating CHeB stars from being tidally spun up, especially for massive CHeB stars. Our prescription may therefore overestimate the number of rapidly spinning secondary BHs, but quantifying this effect will require the implementation of more complicated tidal physics.

\section{GW Detection Prospects}
\label{sec:gw_detections}

\begin{figure}
    \centering
    \includegraphics[width=\linewidth, trim={35 0 0 0}]{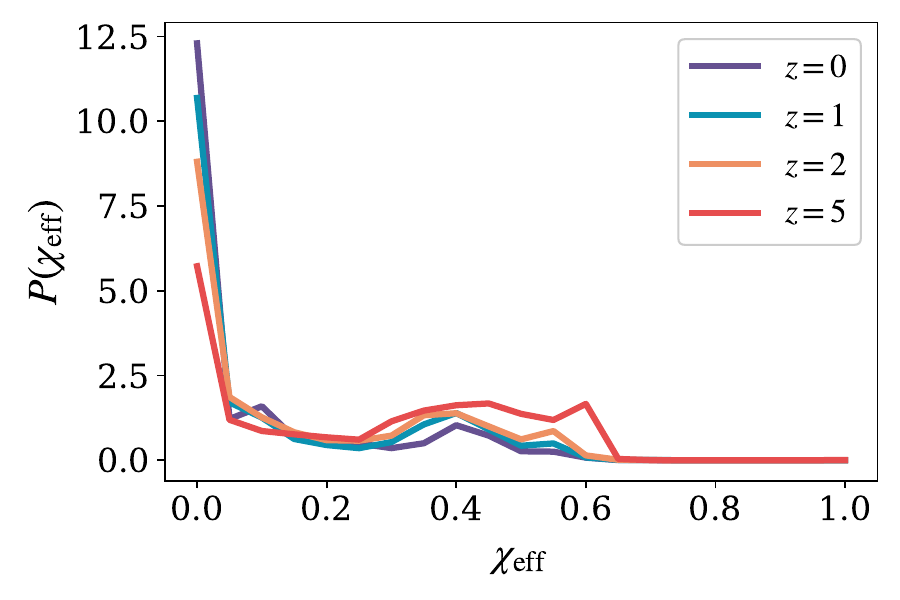}
    \caption{The distributions of $\chi_{\rm eff}$ for binaries at several merger redshifts in the KAPIL26 simulation; all binaries merging at that redshift are included, weighted by their respective merger rates, without accounting for observational selection effects. All distributions are independently normalized.}
    \label{fig:chi_eff_stacked}
\end{figure}

\begin{figure*}
    \centering
    \includegraphics[width=\linewidth]{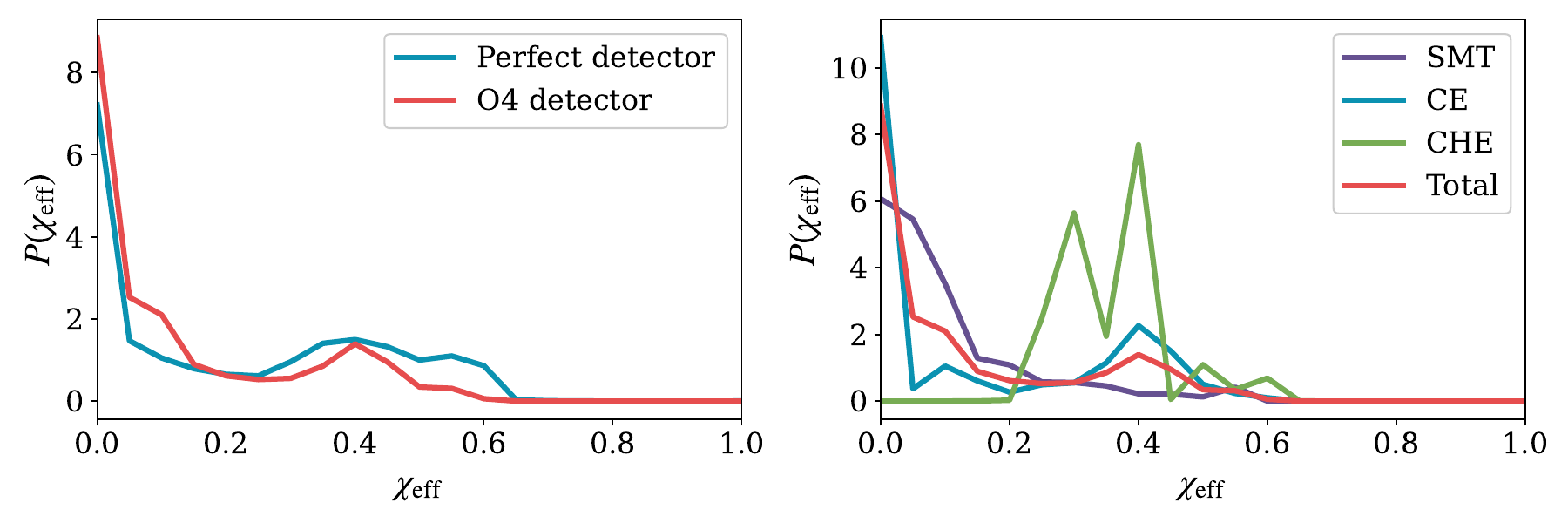}
    \caption{Left panel: the distribution of $\chi_{\rm eff}$ among detectable BBHs from the KAPIL26 population, assuming a typical O4-like sensitivity and the assumption that perfect detectors can detect all mergers in the visible Universe.  Right panel: the distributions of $\chi_{\rm eff}$ among detectable BBHs from the KAPIL26 population assuming a typical O4-like sensitivity, split up by formation channel.  All distributions are independently normalized for visual clarity.}
    \label{fig:3d_subpopulations_det_O4}
\end{figure*}

Not all merging BBHs will be observable, as GW detectors impose selection effects on the observed population.  Broadly, nearby sources and sources containing more massive BHs are louder than distant and/or low-mass sources, and therefore easier to detect.  As a proxy for detectability by the detector network during the O4 run, we consider whether a BBH would have a signal-to-noise ratio above a threshold of 8 in a single instrument, using the noise power spectral density measured in June 2023 in the Livingston detector.  We account for isotropic binary sky locations and orientation and use inspiral-merger-ringdown gravitational waveforms without considering the impact of spin on the signal \citep[see, e.g.,][]{COMPAS_2022}.  While this is a significant over-simplification of the search sensitivity, it allows us to investigate the key impacts of selection effects.  These are shown in Fig.~\ref{fig:3d_subpopulations_det_O4}.

The left panel of Fig.~\ref{fig:3d_subpopulations_det_O4} shows the distribution of $\chi_{\rm eff}$ among BBHs that are observable at the typical sensitivity of O4.  We also show the same distribution for an imaginary detector that would be perfectly sensitive to all BBH mergers in the visible Universe. As detector sensitivity improves through O5 and toward third-generation detectors, such as the Einstein Telescope \citep{Punturo:2010zz, Hild:2010id, ET:2019dnz, ET:2025xjr} and Cosmic Explorer \citep{Reitze:2019iox, Evans:2021gyd}, the observed spin distribution will shift toward this ``perfect'' detector sensitivity.  While more sensitive detectors will be sensitive to mergers at higher redshifts where spins tend to be greater, the evolution is not dramatic, consistent with our expectations from Fig.~\ref{fig:chi_eff_stacked}.  For a perfectly sensitive detector, we expect that more than a third of detected BBHs will have $\chi_{\rm eff}<0.05$ and more than half will have $\chi_{\rm eff}<0.2$.  On the other hand, 15\% will have $\chi_{\rm eff}>0.5$, although only 3\% of detected BBHs at O4 sensitivity will have $\chi_{\rm eff}>0.5$ in our models. 

Recent works have investigated $z-\chi_{\rm eff}$ correlations in the observed BBH catalog \citep[see, e.g.,][]{Biscoveanu:2022qac, LIGOScientific:2025pvj, LIGOScientific:2026ctl}, finding model-dependent evidence for a broadening of the $\chi_{\rm eff}$ distribution with redshift. This is consistent with our simulations, where the inclusion of higher spin BBHs at higher redshifts will widen a redshift-dependent gaussian model. Our simulations also suggest that the mean of the $\chi_{\rm eff}$ distribution should increase with redshift, which has not yet been confidently measured in the current BBH catalog. We expect this feature to become more prominent in third-generation detectors.

/Convolution with the star formation history and detectability impacts the relative weightings of different channels.  For example, while the CHE channel was already rare, contributing only a couple of percent to the BBH yield when integrated across the log-uniform metallicity distribution, this contribution drops to the negligible fraction of less than one billionth after accounting for O4 sensitivity. We do not explore parameter estimation uncertainty in this work, but imperfect measurements of BBH spins \citep[e.g.,][]{LIGOScientific:2025slb, LIGOScientific:2025pvj, Prasad:2026idz} will make it even harder to see substructure or cross-correlations in the spin distribution arising from isolated binary evolution.

\section{Conclusions}
\label{sec:conclusions}

In this work, we have presented BBH spin predictions from isolated binary evolution using a new, self-consistent tidal dissipation implementation (KAPIL26) in COMPAS. We have compared these against predictions from the default COMPAS model without tides (LEGACY), and an idealized instantaneous synchronization and circularization tidal model (PERFECT). Our predicted distributions of $\chi_{\rm eff}$, $M_{\rm tot}$ and $q$ are well-suited for direct comparison against non-parametric and flexible parametric analyses of the growing GW catalog \citep{Tiwari:2020otp, Edelman:2022ydv, Biscoveanu:2022qac, Callister:2023tgi, Heinzel:2024hva, Alvarez-Lopez:2026ymo}.
However, the high-spin tail of the effective spin distribution is depleted by selection effects among BBHs observable with the current detector network, with higher spins becoming accessible as instrument sensitivity grows.
With this caveat in mind, population-level features in $\chi_{\rm eff}$ remain some of the most promising observational probes of tidal dissipation and mass transfer physics in stellar-mass compact binaries. We note some observationally relevant predictions below.

First, we expect the $\chi_{\rm eff}$ distribution in an astrophysically motivated BBH population to be correlated with merger redshift. Specifically, with the KAPIL26 model for tides, we expect the fraction of detectable BBHs with $\chi_{\rm eff}>0.2$ to increase five-fold as detectors improve from O4 to third-generation sensitivities. Note that we only consider the isolated binary evolution scenario in this work, and the redshift-dependence of the overall BBH spin distribution may differ due to the presence of other formation channels.

Comparison against the physically motivated KAPIL26 tidal model demonstrates that the assumption of tidal synchronization of the progenitor of the second-born BH prior to SN does not always hold. When this assumption is made (as in the LEGACY and PERFECT models), tidal models may under- or over-predict the magnitude of $a_2$ by enforcing synchronization in stripped secondaries for which dynamical tides may be relatively weak.

Across all prescriptions and formation channels within isolated binary evolution, almost all first-born BHs form with negligible spin. This prediction holds across the tidal dissipation prescriptions considered in this work, but depends on the efficiency of AM transport in massive stars. The rapidly rotating second-born BH may, however, be the more massive BBH component due to MRR, which occurs for roughly $50\%$ of BBHs merging within 13.8 Gyr in our simulations. Since most of the MRR binaries in our simulations come from the SMT channel, we may be able to identify an isolated SMT BBH from a combination of high primary spin and total mass above $40 M_\odot$. We note that such binaries may also be born dynamically or via hierarchical mergers, which is important to keep in mind when interpreting any observed BBH systems. Additionally, given that high-spin SMT binaries preferentially have shorter coalescence times and merge at higher redshifts, this signature may not be fully accessible until next-generation GW detectors come online.

CHE binaries are notable exceptions when it comes to the spins of first-born BHs from isolated evolution. In our simulations, CHE BBHs occupy a narrow region of equal mass ratios, high total masses, and potentially large spins. This distinctive footprint in the joint $M_{\rm tot}$-$q$-$\chi_{\rm eff}$ space offers a promising avenue for identifying CHE binaries with future GW observations; however, we expect detectable CHE BBHs to be vanishingly rare at current instrument sensitivity, and to contribute less than 1\% of all detections even for third-generation instruments.

Only a select few formation scenarios produce binaries with $\chi_{\rm eff} \gtrsim 0.7$ in our simulations, specifically the CHE channel and the CE channel where both supernovae take place only after the CE. As long as AM transport is efficient in massive stars and our rigid-body assumption is realistic, high values of $\chi_{\rm eff}$ observed at unequal mass ratios could be strong indicators of dynamical assembly or hierarchical mergers \citep{Berti:2008af, Gerosa:2017kvu, Fishbach:2017dwv, Rodriguez:2019huv, Tagawa:2021ofj, Gerosa:2021mno}.

Although our models predict distinct, formation history-dependent features in the spin distribution of merging BBHs, these features are challenging to study in detected populations. Mechanisms such as angular momentum loss during BH formation as well as practical limitations such as GW measurement uncertainty may smooth out the substructure present in the $\chi_{\rm eff}$ distribution of BBHs, making it challenging to learn about tidal theory from spin observations alone. However, as GW catalogs grow with the LVK O4 observing run and beyond, even large-scale features in the spin distribution may be enough to begin disentangling binaries born from the isolated formation scenario from other channels. Furthermore, correlations and features across parameters at the population scale may become powerful tools to enrich our understanding of the formation histories of BBHs. Our work presents a modern, self-consistent prediction of the spin distribution of BBHs from the isolated binary channel using updated tidal theory, which can form the basis of future population modeling efforts.

\section{Acknowledgments}
V.K. is grateful to Bore Gao and Francesco Iacovelli for helpful discussions.
V.K.~and I.M.~acknowledge support from the Australian Research Council (ARC) Centre of Excellence for Gravitational Wave Discovery (OzGrav), through project number CE230100016. 
V.K. and E.B. are supported by NSF Grants No.~AST-2307146, No.~PHY-2513337, No.~PHY-090003, and No.~PHY-20043, by NASA Grant No.~21-ATP21-0010, by John Templeton Foundation Grant No.~62840, by the Simons Foundation [MPS-SIP-00001698, E.B.], by the Simons Foundation International [SFI-MPS-BH-00012593-02], and by Italian Ministry of Foreign Affairs and International Cooperation Grant No.~PGR01167.
E.G. acknowledges support from the ARC Discovery Early Career Research Award (DECRA) DE260101802.
Part of the numerical work was carried out at the Advanced Research Computing at Hopkins (ARCH) core facility (\url{https://www.arch.jhu.edu/}), which is supported by the NSF Grant No.~OAC-1920103.  This research was supported in part by grant NSF PHY-2309135 to the Kavli Institute for Theoretical Physics (KITP) and grant PHY-2210452 to the Aspen Center for Physics.

\newpage

\appendix

\section{Bavera et al. (2021) spin magnitudes}
\label{sec:bavera_spin_mags}

\begin{figure}
    \centering
    \includegraphics[width=\linewidth, trim={35 0 0 0}]{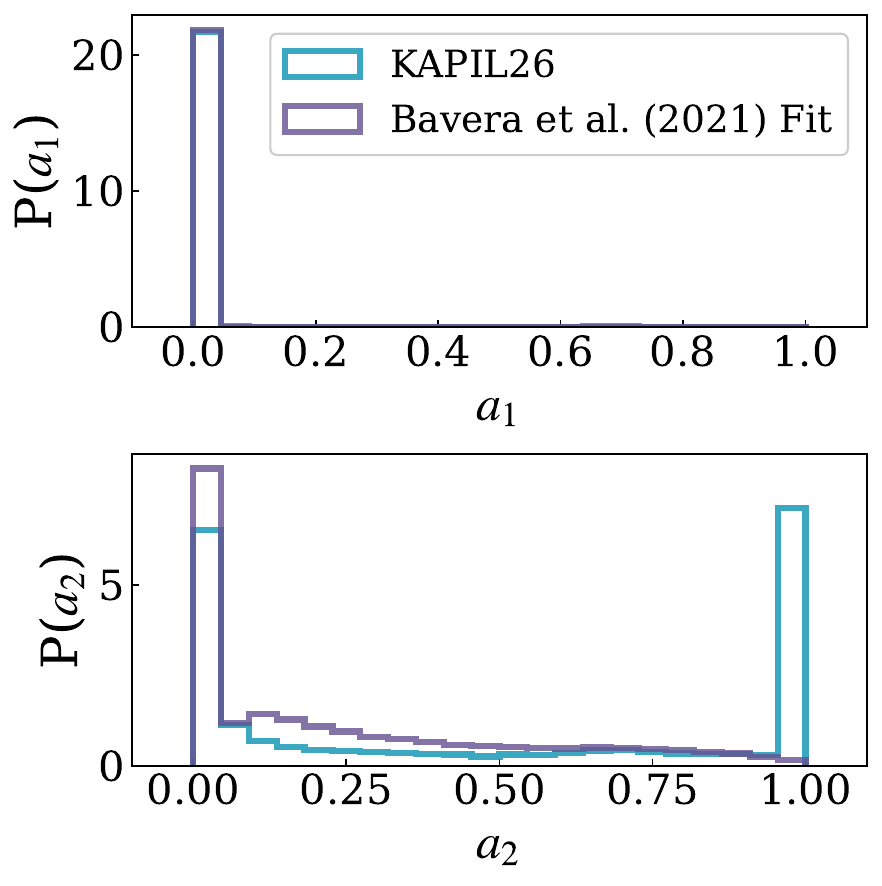}
\caption{Dimensionless BH spin distributions for the primary component at ZAMS (top panel) and the secondary component at ZAMS (bottom panel). The default KAPIL26 distribution is shown in blue, and spins obtained by using the \citet{Bavera:2021evk} fit on the same population are shown in purple.}
    \label{fig:bavera_a1_a2}
\end{figure}

\begin{figure}
    \centering
    \includegraphics[width=\linewidth, trim={20 0 0 0}]{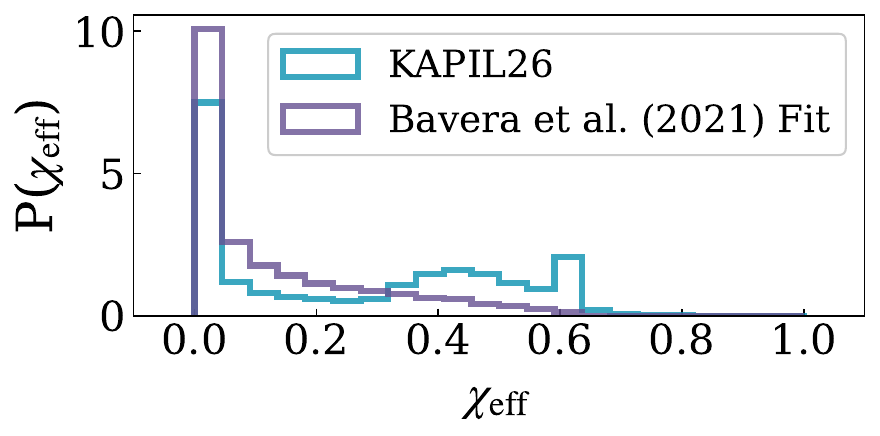}
\caption{$\chi_{\rm eff}$ distributions for the merging BBHs with the KAPIL26 population using the fiducial tidal model (blue), and using the \citet{Bavera:2021evk} fit (purple) to determine spin magnitudes.}
    \label{fig:bavera_chi_eff_breakdown}
\end{figure}

The spin magnitudes of BHs following supernovae in this work depend on our assumption of angular momentum conservation, as described in Sec.~\ref{sec:spin_magnitudes}. However, there are other theoretical frameworks that predict smaller remnant spins at the kinds of orbital separations seen in our simulations. For instance \citet{Bavera:2021evk, Bavera:2020uch} provide a fit to BH spins that emerge from WR-BH binaries below orbital periods of 1 day, which are based on MESA simulations. The assumptions on tidal dissipation made in their model are inconsistent with our simulations, but we present a version of our results that employs their spin magnitude fits anyway.

Only the second-born BH in \citet{Bavera:2021evk} is allowed to have non-zero spin, and the final spin of the secondary is determined by a combination of the orbital period of the WR-BH binary and the mass of the WR star. Our three tidal models produce generally similar orbital separations at the WR-BH stage, and the key differences are in initial angular momenta. To remain as faithful as possible to the \citet{Bavera:2021evk} model, we ignore the angular momenta of the WR stars and simply rely on the orbital properties and WR mass to calculate the remnant spin. Since the results of this calculation are nearly identical across all 3 of our populations, we show only the results from KAPIL26, our fiducial population, in this section. 

In Fig.~\ref{fig:bavera_a1_a2} and Fig.~\ref{fig:bavera_chi_eff_breakdown}, we show the effects of applying the spin magnitude fits from \citet{Bavera:2021evk} to our simulations. The spins of the second-born BH are heavily suppressed, with very few maximally spinning BHs in agreement with \citet{Bavera:2020inc}. The $\chi_{\rm eff} = 0.5$ peak from the default KAPIL26 distribution is also suppressed. This comparison highlights the dominant spin-up mechanism in the default KAPIL26 model: the retention of super-synchronous rotation from prior binary interactions. The \citet{Bavera:2021evk} fit depends only on the orbital period, and does not account for mass transfer history, which is why the second-born BHs have much lower spins.

\section{Comparison to Zahn (1977)}
\label{sec:z77_spins}

\begin{figure*}
    \centering
    \includegraphics[width=\linewidth, trim={35 0 0 0}]{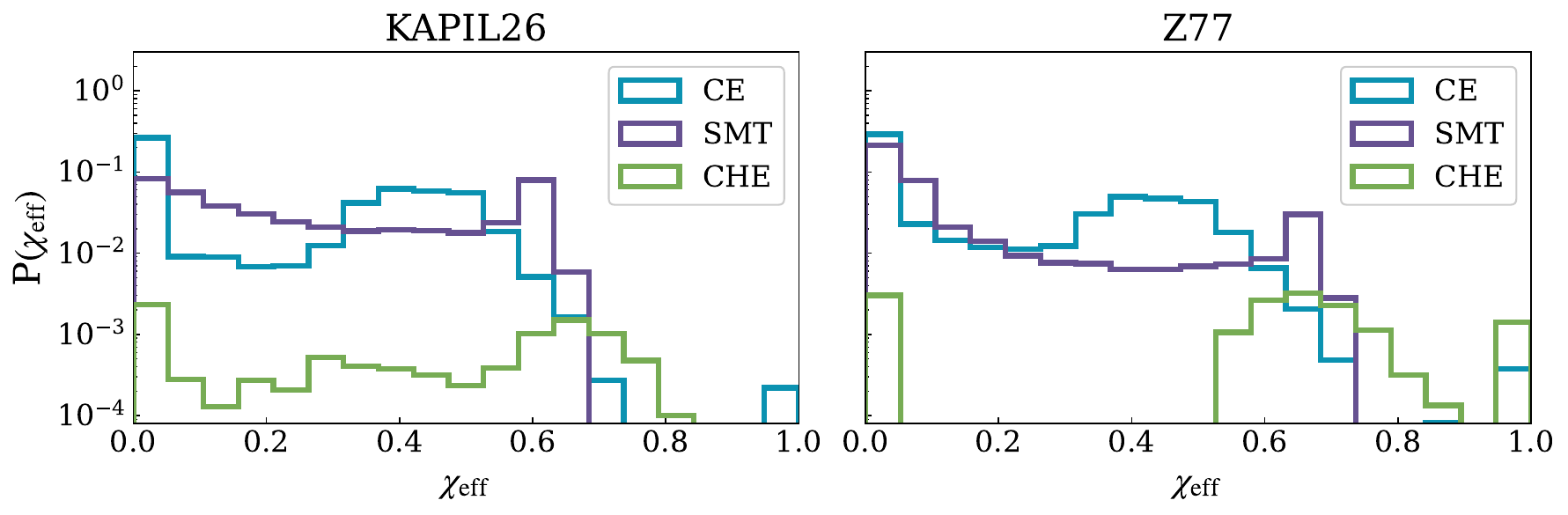}
    \caption{$\chi_{\rm eff}$ probability distributions of merging BBHs with the KAPIL26 (left) and Z77 (right) tides models, split by the CE, SMT, and CHE formation scenarios. The distributions are weighted to reflect the metallicity-specific star formation history of the Universe as per Sec.~\ref{sec:bbh_astro_population}.}
    \label{fig:z77_chi_eff_breakdown}
\end{figure*}

In our companion paper \citep{Kapil:2026hiw}, we present our fiducial tidal model in contrast to the commonly used prescriptions from \citet{zahn_tidal_1977} and \citet{Hurley:2002rf}, which we abbreviate as Z77. For the sake of comparison, we present the outcome of COMPAS simulations performed with the Z77 prescription of tidal dissipation. The full details of our implementation of this model can be found in Appendix B of \citet{Kapil:2026hiw}.

In Fig.~\ref{fig:z77_chi_eff_breakdown} we show the $\chi_{\rm eff}$ distribution for merging BBHs from Z77, with the KAPIL26 result plotted for comparison. The CE spin distributions are very similar between the two tidal models, with mild differences arising for low-spin BHs. The key tidal dissipation mechanism involved in synchronizing the post-CE stripped HeMS and HeHG stars is dynamical IGW dissipation. In the KAPIL26 model, the convective core in HeHG stars is too small for dynamical modes to be excited from the core-envelope boundary, and IGW dissipation is very weak for the widest post-CE binaries. Z77 tides depend only on the overall mass and radius of the star, which does not account for the size of the convective core in COMPAS. Therefore, the strength of dynamical tides for the widest post-CE BH-HeHG binaries is higher in the Z77 model, and the $\chi_{\rm eff} = 0$ peak has a tail that extends to $\chi_{\rm eff} \sim 0.2$. The high-spin CE binaries in the two sets of simulations are very similar, as they have narrow enough post-CE separations to synchronize tidally in KAPIL26 as well as Z77.

The spin distribution for SMT binaries exhibits more prominent differences between Z77 and KAPIL26. Once again, the underlying cause for the differences is the stronger IGW dissipation in Z77 for HeMS and HeHG stars with small core-envelope boundaries. The more efficient tidal coupling between the super-synchronous secondary and the orbit causes the orbit to widen prior to the next MT episode with the Z77 prescription. The wider orbits as well as stronger tidal coupling for evolved stars relative to KAPIL26 lead to a much more pronounced $\chi_{\rm eff}\approx0$ population in the Z77 simulation.

The CHE spin-up in the Z77 model is also affected by the stronger dynamical tides, whereby every CHE BBH remains tidally locked until supernova. This is not true for the KAPIL26 model, where stars can become tidally uncoupled when they evolve off the main sequence, and there is a wider spread of spins.

\bibliography{refs}{}
\bibliographystyle{aasjournalv7}

\end{document}